\documentclass{article}
\usepackage[preprint]{neurips_2025}
\usepackage[utf8]{inputenc} 
\usepackage[T1]{fontenc}    
\usepackage{hyperref}       
\usepackage{url}            
\usepackage{booktabs}       
\usepackage{amsfonts}       
\usepackage{nicefrac}       
\usepackage{microtype}      
\usepackage{xcolor}         
\usepackage{multicol}
\usepackage{multirow}
\usepackage{wrapfig}

\usepackage{amsmath,amsthm,mathtools,amsfonts,amssymb}
\usepackage{graphicx}
\usepackage{algorithm, algorithmic}
\usepackage{subfig}
\usepackage{soul}

\title{Geological and Well prior assisted full waveform inversion using conditional diffusion models}

\author{%
Fu Wang$^{1}$,~Xinquan Huang$^{1}$,~Tariq Alkhalifah$^{1}$\\
$^1$King Abdullah University of Science and Technology\\
\texttt{\{fu.wang,xinquan.huang,tariq.alkhalifah\}@kaust.edu.sa}\\
}

\begin{document}
\maketitle

\title{Geological and Well prior assisted full waveform inversion using conditional diffusion models}

\begin{abstract}
Full waveform inversion (FWI) often faces challenges due to inadequate seismic observations, resulting in band-limited and geologically inaccurate inversion results. 
Incorporating prior information from potential velocity distributions, well-log information, and our geological knowledge and expectations can significantly improve FWI convergence to a realistic model. 
While diffusion-regularized FWI has shown improved performance compared to conventional FWI by incorporating the velocity distribution prior, it can benefit even more by incorporating well-log information and other geological knowledge priors.
To leverage this fact, we propose a geological class and well-information prior-assisted FWI using conditional diffusion models. 
This method seamlessly integrates multi-modal information into FWI, simultaneously achieving data fitting and universal geologic and geophysics prior matching, which is often not achieved with traditional regularization methods.
Specifically, we propose to combine conditional diffusion models with FWI, where we integrate well-log data and geological class conditions into these conditional diffusion models using classifier-free guidance for multi-modal prior matching beyond the original velocity distribution prior. 
Numerical experiments on the OpenFWI datasets and field marine data demonstrate the effectiveness of our method compared to conventional FWI and the unconditional diffusion-regularized FWI. 

\end{abstract}

\section{Introduction}
Our industry has long strived to integrate the seismic value chain, empowering seismic interpreters to play a more significant role in guiding the velocity model-building process. 
Their expertise in defining geological features, such as faults, horizons, and lithology, is fundamental for accurate model construction.
As a result, we have increasingly integrated advanced technology, which can utilize that expertise, into our workflows to enhance the velocity model-building process from seismic data. 
We have developed specialized software that enables interpreters to interactively analyze seismic volumes, mark geological features, and validate the accuracy of inverted models.
However, these approaches are typically post-inversion and aimed at providing guidance for corrections. 
They are often not incorporated within the inversion process itself, as is the case with Full Waveform Inversion (FWI). 
Although FWI can reconstruct the subsurface velocity model by matching the simulated seismograms with the observed recordings, its performance on realistic data is not as good as expected, often due to imperfect observations, resulting in band-limited, noisy, and sparse data.
Incorporating the expertise of geologists and other prior information into FWI is crucial for guiding the inversion toward realistic, high-resolution models.

From the perspective of ill-posed inverse problems, the prior information can be incorporated via regularization.
In addition to standard (fixed) regularization methods, such as Tikhonov \citep{Aster2011} and total variation (TV) regularization \citep{alkhalifah2018full,xiang2016efficient}, which may yield smooth or piece-wise smooth inverted results, respectively, using deep learning to incorporate prior information, like our expectations on geological structures, migrated images, or well information, to guide the inversion has recently gained attention. 
\cite{zhang2018multiparameter} proposed to utilize the facies-based constraints inverted from Bayesian theory to improve multiparameter elastic FWI. 
\cite{singh2018bayesian} also proposed to use facies distribution and the available well logs to derive the prior model as a constraint in the inversion workflow. 
\cite{li2021target} and \cite{zhang2022regularized} proposed the use of a deep neural network to learn the statistical relationship between the inverted model and the facies interpreted from well logs for inversion. 
In addition to these explicit ways, an effective and implicit choice is to use generative models that can build the data distribution, e.g., generative adversarial network (GAN), trained on a velocity distribution based on our expectations to regularize FWI \cite[]{mosser2020stochastic}.
However, we know that it is difficult to train GANs and may face the challenge of model collapse. 
Diffusion models, which show state-of-the-art performance compared to other generative models in generative quality and diversity on image generation tasks \cite[]{dhariwal2021diffusion}, have gained a lot of attention lately.
In addition, the velocity-model-sized latent space in diffusion models enables seamless integration of pre-trained diffusion models (with prior information) into FWI in a plug-and-play manner \citep{wang2023prior}. 
As demonstrated by \cite{wang2023prior}, for FWI, a trained diffusion model can inject prior information of the trained velocity distribution into the inversion results by updating the velocity model in a regularized form after each iteration, yielding the improvement in the accuracy of FWI in case of noisy and sparse observations, referred to as diffusionFWI. 
Although it showed promising performance in field applications, its prior information is guided by the velocity distribution and does not offer a mechanism to inject specific observed prior or high-level expertise from geologists, such as geological preferences and well information, during the inversion.
As mentioned earlier, this information is also crucial to enhance the inversion accuracy and is worth exploring. 
Revisiting the diffusionFWI, it can be regarded as a conditional sampling process with the recorded surface data and the governing equation of wave propagation as conditions. 
Hence, other additional prior information, such as well logs or geological structure information, can also be considered as conditions for the diffusionFWI. 
\cite{wang2024controllable} demonstrated that using a conditional diffusion model can generate velocity models that satisfy the velocity distribution while also matching the given conditions, including class labels, well logs, and structural information. 

To this end, we propose conditional diffusion regularized FWI using conditional diffusion models to inject, for example, geological or well information into the inversion process. 
The proposed method makes the inverted results not only satisfy the observed data and velocity distribution, but also match the additional information like well logs and geological descriptions.
Specifically, we consider the geological label and the well log and its position, which are commonly available in real applications, in the diffusion models' training as conditions.
Then, during the iterative inversion process, these prior information can be implicitly embedded into FWI by applying this conditional diffusion model. 
To the best of our knowledge, this is the first framework that allows the integration of a geologist’s expertise, such as geological labels (e.g., flatness, curvature, and faults), along with regularization from measurements like well log data, something that was not achievable with conventional regularization techniques or deep learning-based methods.
To summarize, our main contributions are as follows:
\begin{itemize}
    \item We propose a novel geological and well-log prior assisted FWI framework using conditional diffusion models, allowing for the integration of geological expertise into FWI.
    \item Through the combination of diffusion FWI and controllable velocity generation using diffusion models, the proposed method shows flexibility in handling multi-modal prior information.
    \item We evaluate the proposed method on synthetic data corresponding to the OpenFWI velocity models and field marine data, and demonstrate the effectiveness of the proposed method.
\end{itemize}

\section{Methodology}
\label{method}
The objective of the proposed method is to integrate velocity distribution priors, observed well logs, and a general description of the subsurface (class label) as examples of information into the FWI process. 
Those three different modalities of information represent the prior knowledge about subsurface velocity distribution, such as velocity range and general structure, measurements besides the observed data, and expert descriptions from the geologist.
In the following subsections, we will first introduce a general conditional diffusion-regularized FWI framework. Then, we will show how to incorporate multi-modal information to control velocity generation and integrate it into diffusion-regulated FWI.

\subsection{Conditional diffusion regularized FWI}
Full waveform inversion has been used for years to estimate a high-solution subsurface velocity model by matching the simulated seismograms with the observed ones. 
For the purpose of suppressing the noise or providing accurate high-wavenumber estimation, the regularization term is added to the FWI, and its objective function is formulated as
\begin{equation}
\label{equ:fwi}
J=\frac{1}{2}\left\|\mathbf{d}_{\mathrm{obs}}-\mathbf{F}(\mathbf{m})\right\|_2^2 + \mathbf{R(m)},
\end{equation}
where $\mathbf{d}_{\mathrm{obs}}$ is the observed data, $\mathbf{m}$ represents the velocity model, $\mathbf{F}$ is a modeling operator (here corresponding to the acoustic wave equation with constant density), and $\mathbf{R}$ denotes a regularization term on the velocity model.
The regularization operator $\mathbf{R}$ can be an explicit regularization term directly applied to the velocity model like Tikhonov regularization \citep{Aster2011}, total variational regularization \citep{alkhalifah2018full,xiang2016efficient} or sparse promoting regularization \citep{zhu2017sparse,huang2019robust}, as well as an implicit operation like a proximal operator in the form of a Plug-and-play regularization \citep{hurault2022proximal}. 

Our goal is to integrate the geologist's expertise and any prior information into the velocity model building. 
Formulated in mathematical terms, that is, given the conditions $\mathbf{c}$ including well log and expected geological priors such as the general description of the subsurface structure, we draw samples from the conditional distribution $p(\mathbf{m}|\mathbf{c})$ as the prior to assisting the velocity estimation of FWI.
Thus, as suggested by \cite{wang2023prior}, the prior is incorporated in a plug-and-play fashion in the inverse problem.
On the topic of building the conditional distribution, we have many options, such as variational autoencoder \citep{kingma_auto-encoding_2013}, generative adversarial networks \citep{goodfellow_generative_2014}, normalizing flows \citep{rezende_variational_2015}. 
Considering the high-sample-quality ability of diffusion models compared to those generative models and the feasibility of combining the FWI iteration with the diffusion step, we choose the denoising diffusion probabilistic models (DDPMs) to store the conditional distribution of the velocity models $p(\boldsymbol{m}|\mathbf{c})$ in which they were trained on. 

The nature of the DDPMs is to learn a diffusion denoising network, yielding an ability to iteratively transform the sample from a Gaussian distribution to a sample representing the target velocity distribution. 
After substituting the proximal operator of $\mathbf{R}$ with the diffusion denoising network, the iterative update of the velocity model follows this formula:
\begin{equation}
\begin{split}
\mathbf{m}_{k+1}&=f\left(\mathbf{m}_k+\alpha_k \frac{\partial J}{\partial \mathbf{m}_k}, s_{\boldsymbol{\theta}^*}, \mathbf{c},  t\right) +\sqrt{\beta_t} \epsilon,
\end{split}
\label{equ:proxi}
\end{equation}
where $f$ is the reverse diffusion step of the diffusion model, $s_{\boldsymbol{\theta}}^*$ is the diffusion denoising network or the learned score function \citep{song2020score}, $\beta_t$ is the noise schedule at time step, $t$, of the diffusion process, $\epsilon$ is random noise, $\mathbf{m}_k$ is the velocity model at the $k$th iteration, and $\alpha_k$ is the update step length. 
The FWI update iteration is embedded into the diffusion model's generation progress (sampling), which we call condtional diffusion-regularized FWI (or conditional diffusionFWI).
To improve the stability of the inner FWI iteration, the velocity used in the FWI iteration step is free of noise while the noise is added after the FWI update.
The details of $f$ and $s_\theta^*$ will be introduced in the next subsections.

\subsection{Training and inference of diffusion models}
The basic pipeline of the DDPMs includes two important diffusion processes: the forward diffusion process for training and the reverse diffusion process for sampling, a.k. inference.
For the former part, we gradually transform the target velocity distribution to the given described distribution, e.g., the Gaussian distribution for simplicity, and the transition distribution is also given by the Gaussian distribution, written as follows:
\begin{equation}
    \label{equ:xt}
    \mathbf{x}_t = \sqrt{\alpha_t}\mathbf{x}_{t-1} + \sqrt{1-\alpha_t} \mathbf{\epsilon}_{t-1},
\end{equation}
where $\mathbf{\epsilon}$ is the random noise, $\alpha_t=1-\beta_t$, and $\beta_t$ is a noise scheduler to ensure that after T steps, the $\mathbf{x}_t$ satisfies the described Gaussian distribution $\mathcal{N}(0, \mathbf{I})$. 
Thus, given any sampled clean data (seismic velocity models) $\mathbf{x}$ from a given target distribution $p(\mathbf{x})$, the corrupted data at time step $t$ is 
\begin{equation}
\label{forward_p}
    \boldsymbol{x}_t=\sqrt{\bar{\alpha}_t} \boldsymbol{x}_0+\sqrt{1-\bar{\alpha}_t} \boldsymbol{\epsilon}, \boldsymbol{\epsilon} \sim \mathcal{N}(0, \boldsymbol{I}), i=1,\dots, T,
\end{equation}
where $\bar{\alpha}_t=\prod_{j=1}^t (1-\beta_j)$, and $\beta_t$ is the cosine noise scheduler suggested by \cite{nichol_improved_2021}.
As proven in \cite{anderson1982reverse}, the above forward diffusion process can be reversed with a given score function, $\nabla_{\mathbf{x}_t} \log p\left(\mathbf{x}_t\right)$, which is our learning target $s_\theta^*$, by means of the reverse diffusion process 
\begin{equation}
\label{reverse}
    \boldsymbol{x}_{t-1}=\frac{1}{\sqrt{\alpha_t}}\left(\boldsymbol{x}_t+\beta_t s_\theta^*\left(\boldsymbol{x}_t,t)\right)\right)+\sqrt{\beta_t}\boldsymbol{\epsilon}, \quad \boldsymbol{\epsilon} \sim \mathcal{N}(0, \boldsymbol{I}).
\end{equation}
Given a sample from a random distribution, this reverse diffusion process can iteratively remove the noise and transform the image into a clean sample from the distribution $p(\mathbf{x})$.
To obtain a denoising diffusion network $s_\theta^*$, in practice, we randomly sample data $\{x_{i}\}_{i=1}^N$ from the training dataset and obtain the corrupted data using equation~\ref{forward_p}. Note that for each sample, the given time $t$ is also random. 
Then, given a neural network $\Phi_\theta$ with input $\{x_{t_i}\}_{i=1}^N$ and time step $\{t_i\}_{i=1}^N$, we measure the distance between the denoised results and original clean data by
\begin{equation}
    \label{loss}
    \mathcal{L} = \sum_{i=1}^N\|\Phi_\theta(x_{t_i},t_i)-x_{i}\|_2^2
\end{equation}
as the loss function to optimize the parameters of the neural network. As suggested in \cite{ho2020denoising}, we use a residual type neural network $\boldsymbol{\epsilon}_\theta$ to predict the noise given the corrupted noise instead of the direct prediction of the original clean data, yielding $\Phi_\theta(x_{t_i},t_i)=\left(\frac{x_{t_i}}{\sqrt{\bar{\alpha}_{t_i}}}-\sqrt{\frac{1-\bar{\alpha}_{t_i}}{\bar{\alpha}_{t_i}}} \boldsymbol{\epsilon}_\theta\left({x}_{t_i},t_i\right)\right)$.
Substituting this form into equation~\ref{reverse}, the reverse diffusion process is modified to
\begin{equation}
\label{reverse_noise}
    \boldsymbol{x}_{t-1}=\frac{1}{\sqrt{\alpha_t}}\left(\boldsymbol{x}_t-\frac{\beta_t}{\sqrt{1-\bar{\alpha}_t}} \boldsymbol{\epsilon}_\theta\left(\boldsymbol{x}_t, t\right)\right)+\sqrt{\frac{\beta_t(\alpha_t-\bar{\alpha}_t)}{\alpha_t(1-\bar{\alpha}_t)}}\boldsymbol{\epsilon}, \quad \boldsymbol{\epsilon} \sim \mathcal{N}(0, \boldsymbol{I}).
\end{equation}
Thus, we define the loss function used in this paper for noise prediction, which is 
\begin{equation}
    \label{loss_epsilon}
    \mathcal{L} = \sum_{i=1}^N\|\boldsymbol{\epsilon}_\theta(x_{t_i},t_i)-\epsilon_{i}\|_2^2.
\end{equation}
Comparing equations~\ref{reverse} and \ref{reverse_noise}, we found that the learned noise prediction $\boldsymbol{\epsilon}_\theta$ equals to a scaled score function. 

The conditional generation process of the diffusion model can be formulated in the context that given the conditioning signal $\boldsymbol{c}$, which can represent a class label (related to geological information),  well information,  or any other type of condition, the generated output $\boldsymbol{m}$ satisfies the distribution of $p(\boldsymbol{m})$ while consistent with the given condition \citep{wang2024controllable}.
Here, we use a classifier-free guidance diffusion model \citep{ho2022classifier} to include the condition in the training.
Compared to the previously unconditional diffusion model, the only modification is to equip the data sample $\mathbf{m}$ with the condition $\mathbf{c}$, and replace $\boldsymbol{\epsilon}_\theta(\boldsymbol{x}, t)$ to $\boldsymbol{\epsilon}_\theta(\boldsymbol{x},\boldsymbol{c}, t)$. To improve the sample quality, rather than using the prediction $\boldsymbol{\epsilon}_\theta(\boldsymbol{x},\boldsymbol{c}, t)$, we select the following weighting form,
\begin{equation}
\label{noise_predict}
    \hat{\boldsymbol{\epsilon}}_\theta(\boldsymbol{x}_t,\boldsymbol{c},t) = (1+\lambda)\boldsymbol{\epsilon}_\theta(\boldsymbol{x}_t,\boldsymbol{c},t) - \lambda \boldsymbol{\epsilon}_\theta(\boldsymbol{x}_t, t),
\end{equation}
where $\lambda$ is the conditioning scale to adjust the strength of the conditioning. 
In this study, unconditional generation is performed with $\lambda=-1$, while conditional generation uses $\lambda$ values ranging from 1.0 to 20.0.
A higher value for $\lambda$ will decrease the diversity of the samples but increase the consistency between the higher quality (high-fidelity) sample and the given condition \citep{ho2022classifier}. 
During training, to avoid separate training of two neural networks, we apply the classifier-free guidance strategy by randomly dropping the conditional inputs with a probability of 0.5. 
Specifically, for each training sample, the conditioning information is either retained or replaced, with equal probability, with a learnable null embedding initialized from random parameters. 
This enables the model to learn both conditional and unconditional generation behaviors, which can later be combined at inference time for guided sampling.
With this trick in the training stage, the unconditioned $\boldsymbol{\epsilon}_\theta(\boldsymbol{x}_i)$ can be realized by setting $\boldsymbol{c}$ as a vector of null condition in $\boldsymbol{\epsilon}_\theta(\boldsymbol{x}_i,\boldsymbol{c})$.
\begin{figure}
    \centering
    \includegraphics[width=0.7\textwidth]{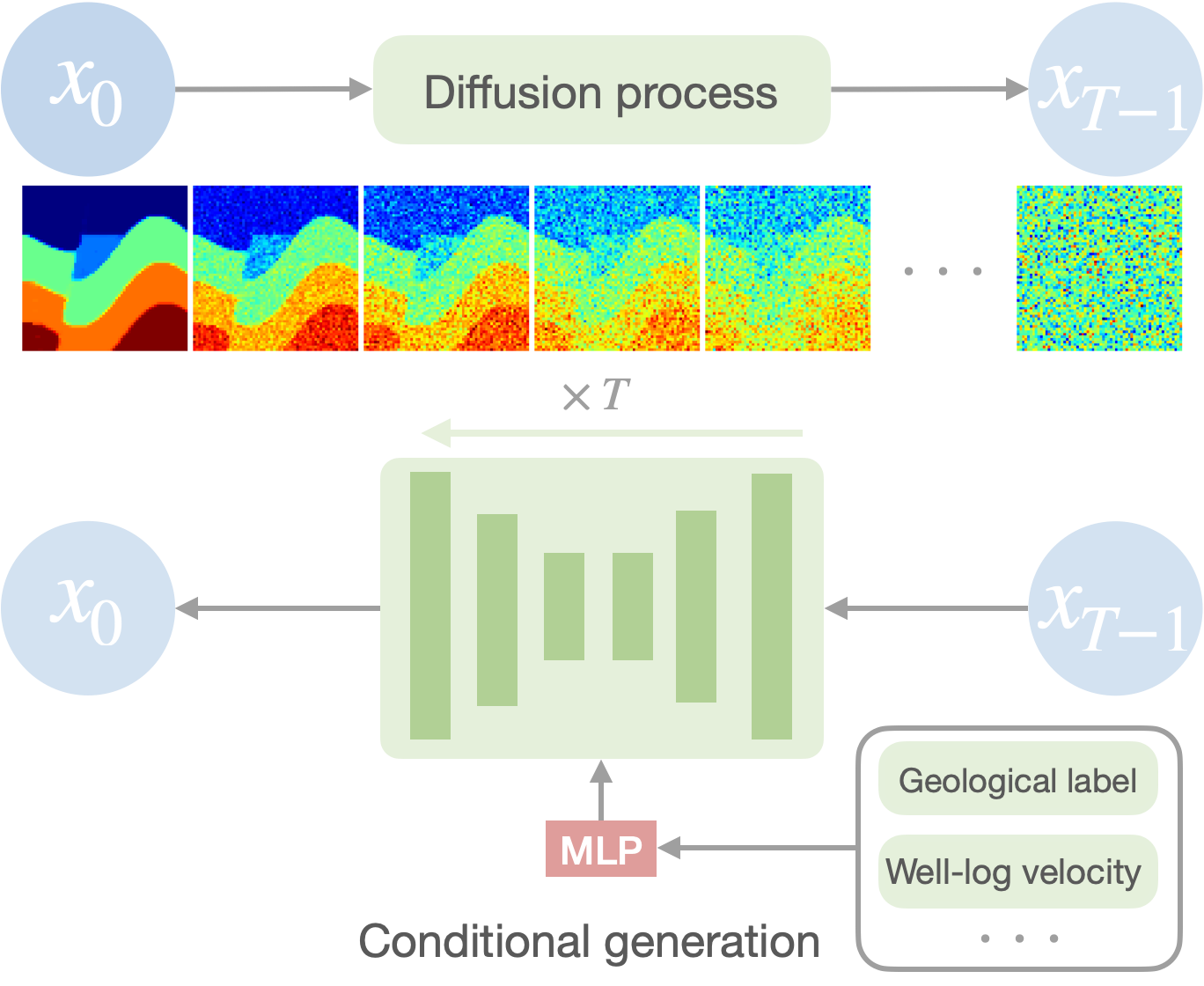}
    \caption{Conditional generative diffusion models where different conditions, such as geological labels and well-log velocities, are incorporated for velocity prediction. The backbone network is U-Net \citep{ronneberger2015u}.$\boldsymbol{x}_0$ represents samples from the training models distribution, which are transformed using the Diffusion process to $\boldsymbol{x}_{T-1}$, which are samples corresponding to a normal distribution, and $\boldsymbol{T}$ is the number of time steps. $\bold{MLP}$ stands for Multi-Layer Perceptron.}
    \label{fig:digram2}
\end{figure}

The total pipeline for training and sampling of the conditional diffusion models is shown in Figure~\ref{fig:digram2}.
During the training, we pair the samples $\mathbf{x}$ and corresponding conditions $\mathbf{c}$, in which we randomly set the condition of some samples to null conditions and corrupt the samples $\mathbf{x}$ with random Gaussian noise using equation~\ref{forward_p} (top in the Figure~\ref{fig:digram2}). Then, we use the neural network to estimate the noise given corrupted samples $\mathbf{x}_t$ and conditions $\mathbf{c}$.
After training, the sampling process (bottom in the Figure~\ref{fig:digram2}) with modified noise prediction (equation~\ref{noise_predict}) is
\begin{equation}
\label{reverse_noise_conditional}
    \boldsymbol{x}_{t-1}=\frac{1}{\sqrt{\alpha_t}}\left(\boldsymbol{x}_t-\frac{\beta_t}{\sqrt{1-\bar{\alpha}_t}} \hat{\boldsymbol{\epsilon}}_\theta(\boldsymbol{x}_t,\boldsymbol{c},t)\right)+\sqrt{\frac{\beta_t(\alpha_t-\bar{\alpha}_t)}{\alpha_t(1-\bar{\alpha}_t)}}\boldsymbol{\epsilon}=f(\boldsymbol{x}_t, \boldsymbol{\epsilon}_\theta,\boldsymbol{c}, t)+\sqrt{\frac{\beta_t(\alpha_t-\bar{\alpha}_t)}{\alpha_t(1-\bar{\alpha}_t)}}\boldsymbol{\epsilon}.
\end{equation}
The backbone network for this noise prediction is U-Net. 
The condition $\mathbf{c}$ is embedded as a vector by a shallow neural network and then added to the hidden layer of the U-Net. 

\subsection{The pipeline of Conditional diffusionFWI}
\begin{figure}
    \centering
    \includegraphics[width=0.6\textwidth]{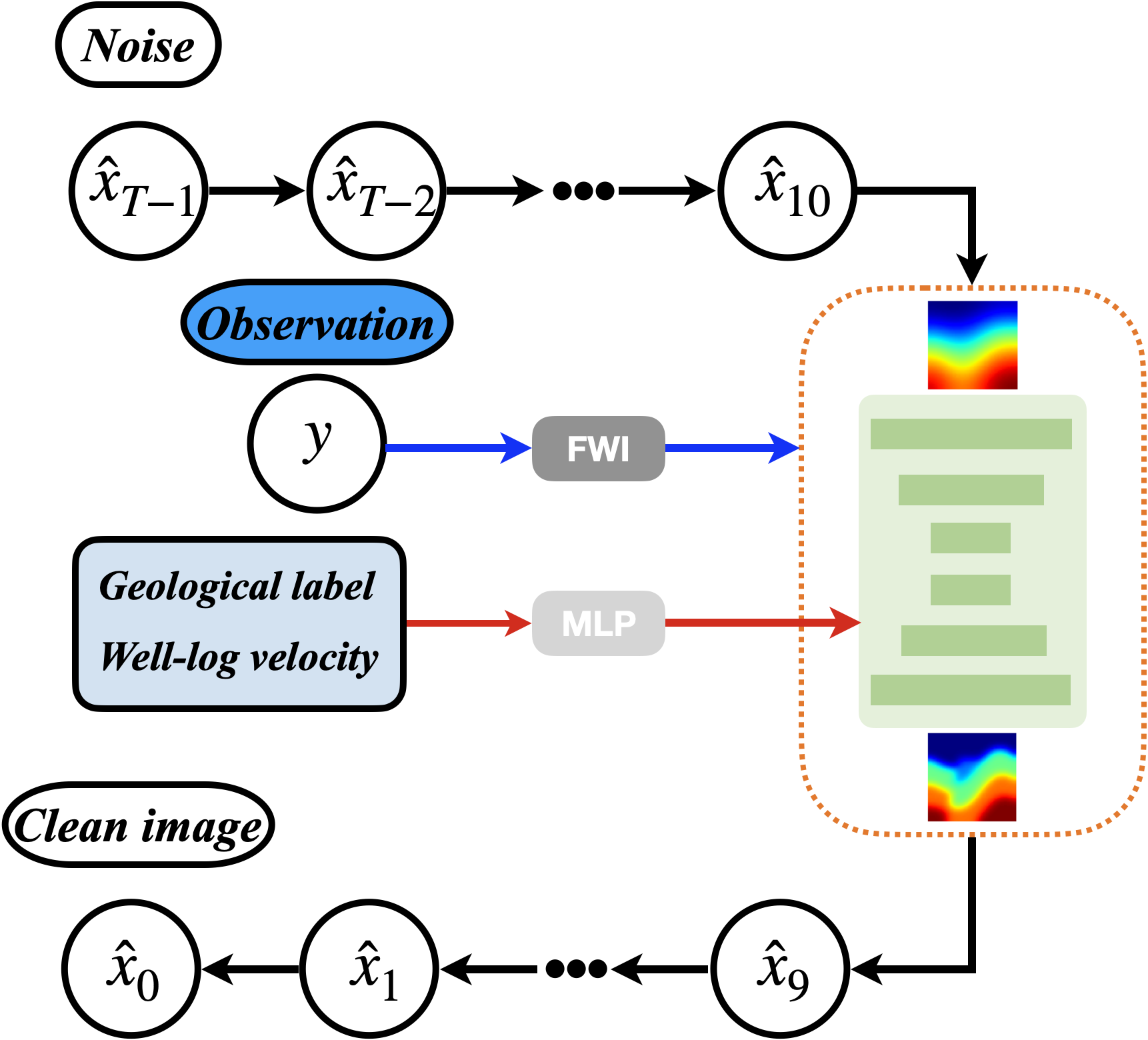}
    \caption{The workflow of our conditional diffusion FWI, including the inner FWI iterations and outer reverse conditional generation progress. The blue arrow denotes the FWI contribution from observation data. The red arrow denotes the reverse conditional generation with prior conditions from well-log velocity and geological labels. The whole workflow includes several reverse diffusion steps. Here, for illustration, we mainly show our method within one reverse diffusion step. However, this is repeated in the other reverse diffusion steps until we approach $\hat{\boldsymbol{x}}_{0}$.}
    \label{fig:ConditionDiffusionFWI}
\end{figure}

Using the reverse diffusion step of the diffusion model $f$ and the learned scaled score function $s_{\boldsymbol{\theta}}^*$ in the equation~\ref{equ:proxi}, we obtain the update of the conditional priors regularized diffusion FWI given by the following form
\begin{equation}
\begin{split}
\mathbf{m}_{k+1}&=\frac{1}{\sqrt{\alpha_k}}\left(\mathbf{m}_k+\alpha_k \frac{\partial J}{\partial \mathbf{m}_k}-\frac{\beta_k}{\sqrt{1-\bar{\alpha}_k}} \hat{\boldsymbol{\epsilon}}_\theta(\mathbf{m}_k+\alpha_k \frac{\partial J}{\partial \mathbf{m}_k},\boldsymbol{c},k)\right)+\sqrt{\frac{\beta_k(\alpha_k-\bar{\alpha}_k)}{\alpha_k(1-\bar{\alpha}_k)}}\boldsymbol{\epsilon}.
\end{split}
\label{equ:proxi2}
\end{equation}
The workflow of the conditional diffusion FWI is the same as diffusion FWI \citep{wang2023prior} except for the conditional input to the backbone U-Net in the diffusion model, as shown in Figure \ref{fig:ConditionDiffusionFWI}.
In practice, we often do several inner regular FWI updates and one outer reverse conditional diffusion step. 
The proposed method is a universal prior regularized FWI, in which any multi-modal information can be incorporated via conditions on the diffusion model. This paper takes geological labels and well log information as additional prior to demonstrate the effectiveness of the proposed framework.

\section{Numerical experiments}
\label{numerical}
In this section, we will demonstrate the effectiveness of the proposed method in handling single and multiple priors using the OpenFWI dataset \cite[]{deng2021openfwi}.
We include the geological class label and the well-log information as typical conditions during the training for the diffusion model and then apply it to the proposed conditional diffusion regularized FWI.
Prior to FWI, we need to train the diffusion model first. 

\subsection{The training of Conditional diffusion models on OpenFWI dataset}
\begin{figure}
    \centering
    \includegraphics[width=0.75\textwidth]{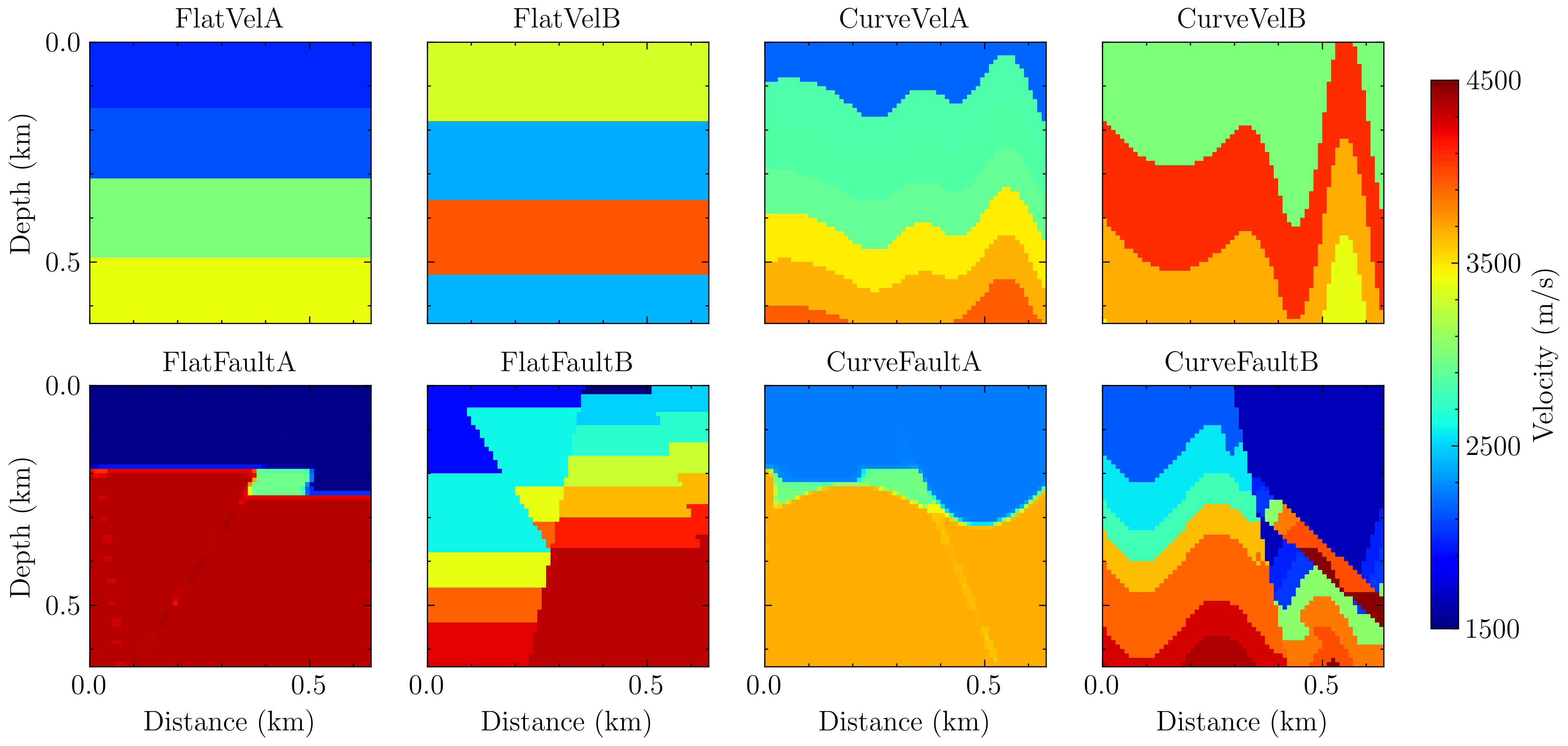}
    \caption{Samples of velocity models corresponding to the various classes of the OpenFWI dataset.}
    \label{fig:samples}
\end{figure}

Here, we train the conditional diffusion model on the OpenFWI dataset to showcase the flexibility of the proposed method to control the velocity generation, specifically with class labels and well logs. 
The dataset comprises eight distinct classes: "FlatVel-A, "FlatVel-B", "CurveVel-A", "CurveVel-B", "FlatFault-A", "FlatFault-B", "CurveFault-A", "CurveFault-B", whose example samples are shown in Figure \ref{fig:samples}. 
They exhibit distinct structural features and velocity variations with depth. 
For example, the class name "FlatVel-" represents layered velocity models, while "CurveVel-" corresponds to velocity models with folding layers. For those two types of velocity models, "A" represents the gradual increase of velocity with depth, while "B" represents random velocity variation with depth.
The "FlatFault-" and "CurveFault-" denote the types of velocity models that include faults with horizontal or curved interfaces, respectively.
For "FlatFault-" and "CurveFault-", "A" indicates the case with fewer discontinuities, while "B" indicates the case with more discontinuities and more pronounced velocity changes.
The original size of the velocity model is 70$\times$70, and we crop it to a model size of 64$\times$64 for ease of training. 
To handle the distribution of this dataset, we use a standard U-Net \citep[]{oktay2018attention} with attention blocks, and the feature maps resolution of the U-Net is given by $[64\times64, 32\times32, 16\times16, 8\times8]$. 
We train the diffusion model using an Adam optimizer with a learning rate of 1e-4. The batch size for the training is 1024, and the maximum training iteration is set to 200000. 
For stable convergence, we also apply the Exponential Moving Average (EMA) for model parameter updates.
The velocity model class label and the well log (a randomly chosen vertical profile from the velocity model) are passed through a multi-layer perceptron (MLP) and then fused directly into the U-Net layers. 
For the well, we incorporate a positional encoding so that the network learns the position of the well. 

\subsection{Synthetic inversion test}

\begin{figure}[htb!]
    \centering
    \subfloat[]{\includegraphics[width=0.32\textwidth]{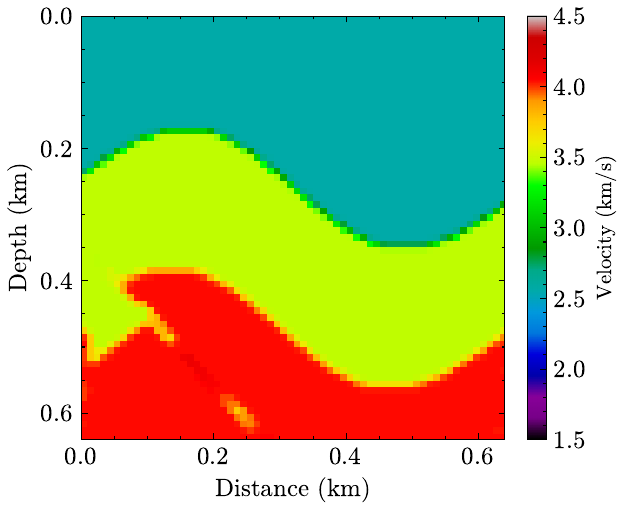}}\label{fig:model_true}
    \subfloat[]{\includegraphics[width=0.32\textwidth]{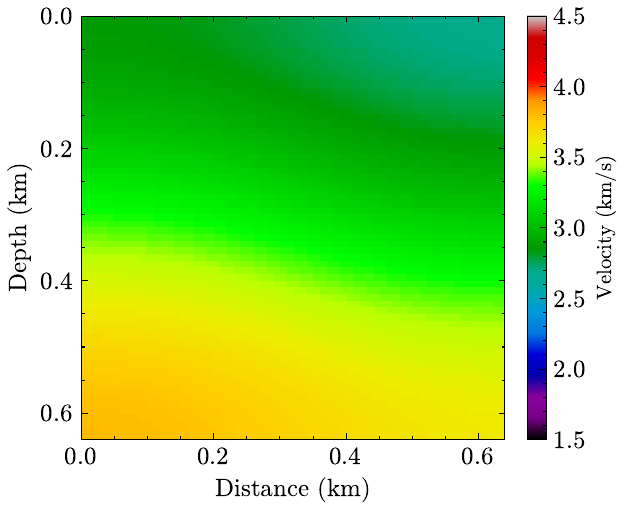}}\label{fig:model_init}\\
    \subfloat[]{\includegraphics[width=0.32\textwidth]{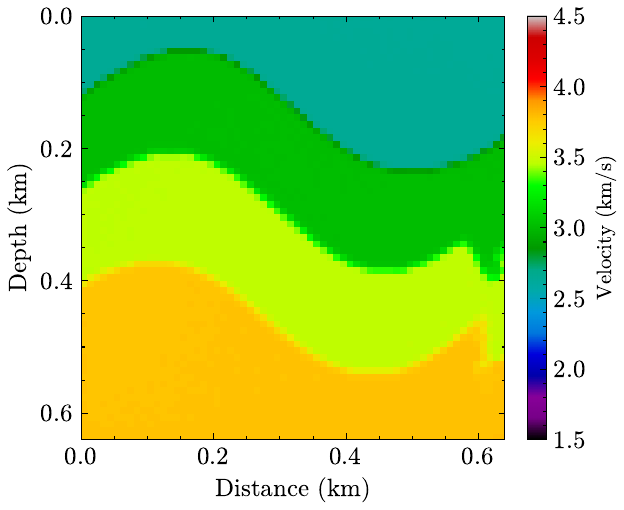}}\label{fig:model_init_class}
    \subfloat[]{\includegraphics[width=0.32\textwidth]{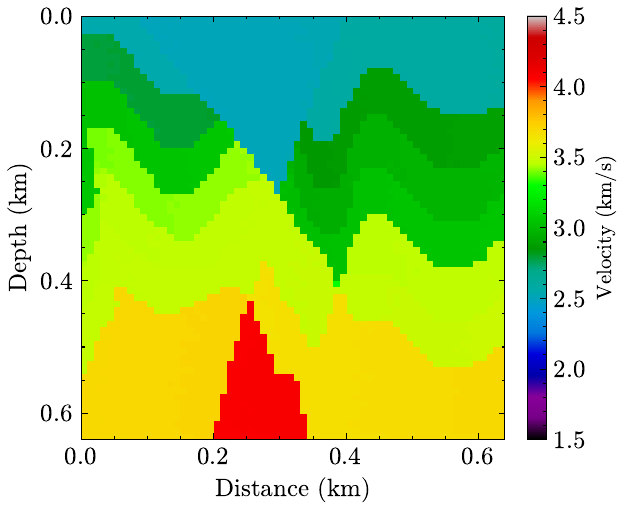}}\label{fig:model_init_well25}
    \subfloat[]{\includegraphics[width=0.32\textwidth]{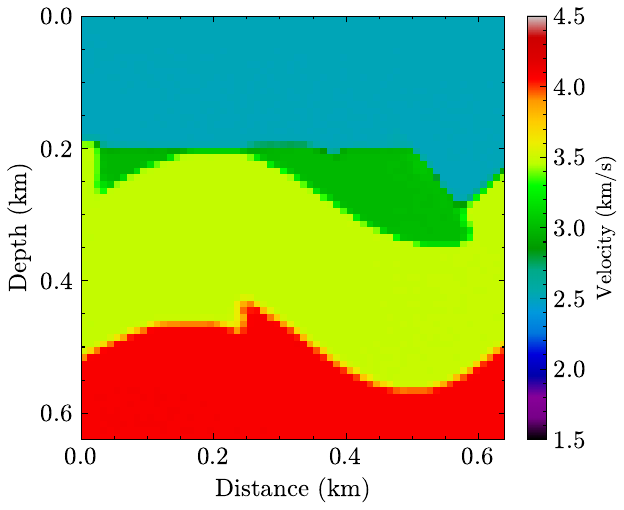}}\label{fig:model_init_class_well25}
    \caption{ a)The true velocity model, and b) the initial velocity model. The pure conditional generation results starting from the initial model using c) the geological class label, d) a well located at 0.25 km, and e) the geological label and the well information together. Noted that there are no FWI incorporated for those generations. This demonstrates the importance of incorporating the observed data by FWI.}
    \label{fig:true&init}
\end{figure}

After the conditional diffusion model is trained, we can use it directly in FWI with practically negligible additional cost to the inversion. 
To illustrate the influence of geological and well-log conditions on diffusion FWI, we select a velocity model from class "CurveFaultA". The true and initial velocity models for this class are shown in Figures~\ref{fig:true&init} (a) and (b). 
This sample was not used in the training (a test sample). 
The size of the velocity model is 64$\times$64 with a 10 m spatial interval in both $x$ and $z$ directions.
Before FWI, we use the pre-trained diffusion model to predict velocity directly starting from the initial model, given the conditions of the class label, well log, or class label and well log, first without any guidance from FWI. 
As shown in Figure~\ref{fig:true&init}(c-e), we can observe the importance of the condition information in the generation, especially when both conditions are used, but the generated models have clear differences with the true model, absent the data fitting guidance. 

\begin{figure}[htb!]
    \centering
    \subfloat[]{\includegraphics[width=0.32\textwidth]{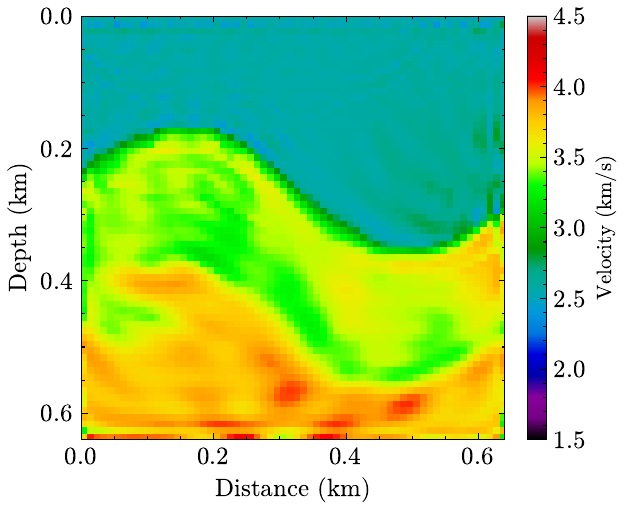}}\label{fig:middle_shot_conv}
    \subfloat[]{\includegraphics[width=0.32\textwidth]{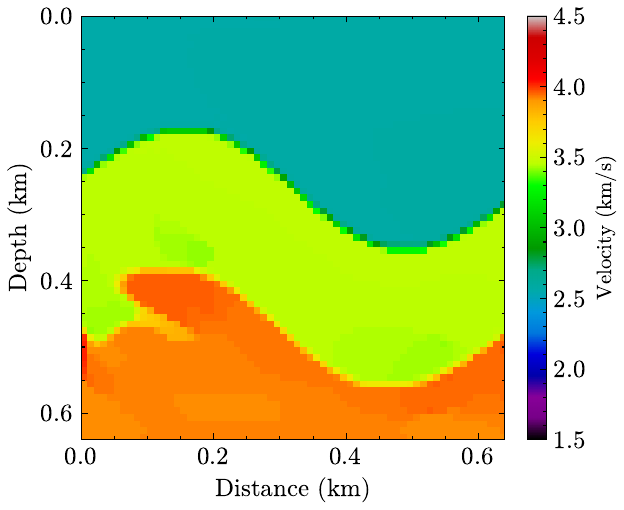}}\label{fig:middle_shot_diffusion}
    \subfloat[]{\includegraphics[width=0.32\textwidth]{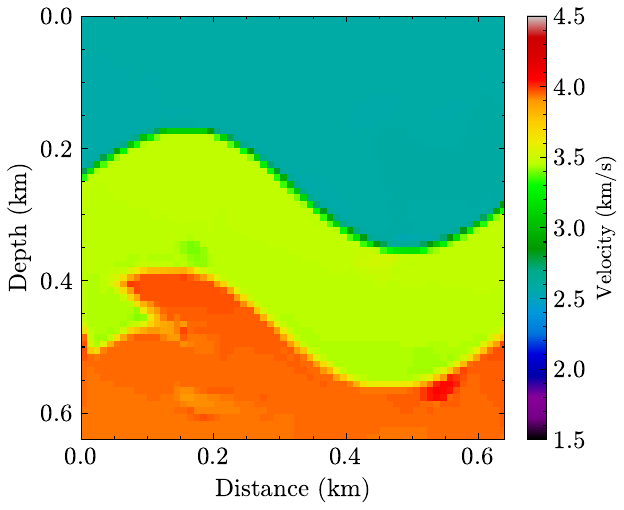}}\label{fig:middle_shot_diffusion_class}
    \subfloat[]{\includegraphics[width=0.32\textwidth]{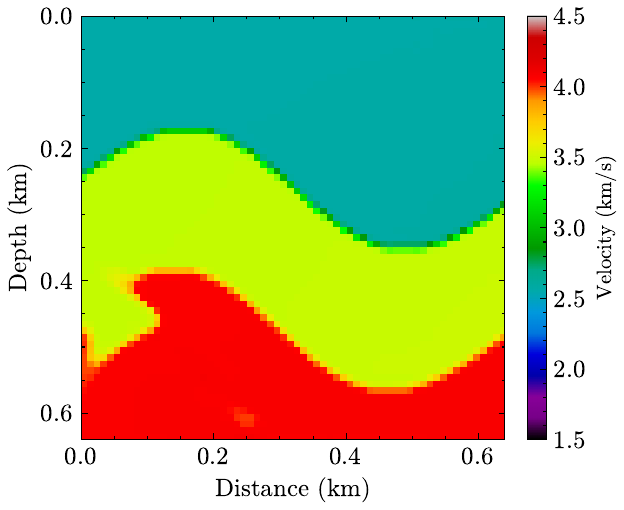}}\label{fig:middle_shot_diffusion_well}
    \subfloat[]{\includegraphics[width=0.32\textwidth]{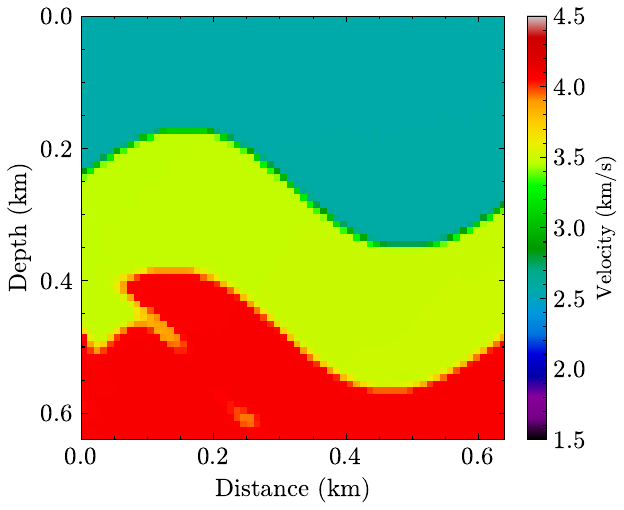}}\label{fig:middle_shot_diffusion_well_class}
    \caption{Inversion results from a) conventional FWI, b) diffusion FWI, and conditional diffusion FWI with c) the geological class label, d) a well located at 0.25 km, and e) the geological class label and the well information together. The proposed conditional diffusion FWI outperforms both conventional methods and standard diffusion-based FWI. Moreover, incorporating more conditioning information leads to improved inversion results.}
    \label{fig:middle_shot}
\end{figure}

\begin{figure}
    \centering
    \includegraphics[width=0.75\textwidth]{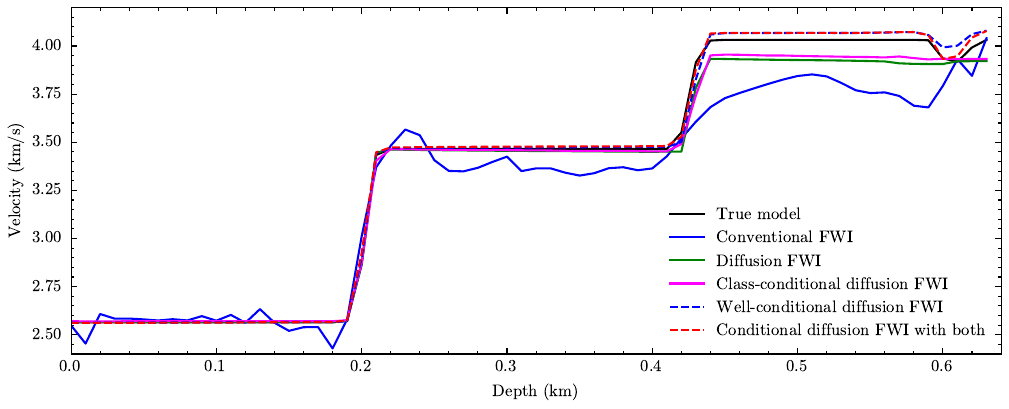}
    \caption{A profile comparison at the location of the well at 0.25 km for the inversion results in Figure \ref{fig:middle_shot}. Detailed profile comparisons further validate that the proposed conditional diffusion FWI outperforms other approaches. Incorporating well constraints improves the match between the inverted results and well data, while jointly using both well and class conditioning yields the best performance.}
    \label{fig:m_profile}
\end{figure}

We compute synthetic observed data by numerically solving the acoustic wave equation with a time sampling interval of 1 ms using a 15 Hz Ricker wavelet with a total recording length of 1.5 s. 
We uniformly place 16 receivers and 4 shots on the surface.  Considering the relatively high-frequency and limited acquisition, applying conventional FWI admits poor results (Figure \ref{fig:middle_shot}a).
Meanwhile, including the diffusion model introduces high-resolution components, as shown in Figure \ref{fig:middle_shot}b. 
As the class label "CurveFaultA" is associated with high-velocity values deep, as well as sinusoidal-shaped curves and fault structures, the velocity values deep are further improved (Figure \ref{fig:middle_shot}c) courtesy of including the geological class. Alternatively, using the well-log condition, as shown in Figure \ref{fig:middle_shot}d, helps improve the velocity values at depth even more. Using both the geological class label and well condition together (Figure \ref{fig:middle_shot}e), the small fault structure is more visible, which can not be achieved by the conventional FWI or regular diffusion FWI. Figure \ref{fig:m_profile} shows the profile comparison at the well location at 0.25 km. We can see that the well condition reasonably constrains the model. 
Table~\ref{tab:model_comparison} reports the quantitative performance of different methods in terms of Structural Similarity Index Measure (SSIM), Peak Signal-to-Noise Ratio (PSNR), and relative L2 error, where “Gen” denotes pure generation without inversion, and “CD-FWI” denotes Conditional DiffusionFWI. SSIM evaluates the perceptual similarity between the predicted and reference images, PSNR reflects the reconstruction fidelity in decibels, and the relative L2 error measures the normalized discrepancy in Euclidean norm. 
It further demonstrates that the more conditions we have, including data, class, and well, the better results we can obtain.
\begin{table}[h]
\centering
\caption{Quantitative comparison of different approaches using PSNR, SSIM, and Relative L2 error. Higher PSNR and SSIM, lower relative L2 error are better.}
\label{tab:model_comparison}
\begin{tabular}{lccc}
\toprule
\textbf{Approaches} & \textbf{PSNR (dB)}$\uparrow$  & \textbf{SSIM}$\uparrow$ & \textbf{Relative L2 error}$\downarrow$ \\
\midrule
\texttt{Gen (class)}            & 23.47 & 0.289 & 0.0772 \\
\texttt{Gen (well)}             & 23.44 & 0.321 & 0.0835 \\
\texttt{Gen (class\&well)}      & 25.49 & 0.495 & 0.0659 \\
\texttt{Conventional FWI}                   & 28.10 & 0.670 & 0.0489 \\
\texttt{Diffusion FWI}                  & 36.49 & 0.900 & 0.0183 \\
\texttt{CD-FWI (class)}       & 37.96 & 0.958 & 0.0156 \\
\texttt{CD-FWI (well)}        & 41.22 & 0.953 & 0.0108 \\
\texttt{CD-FWI (class\&well)} & \textbf{41.51} & \textbf{0.976} & \textbf{0.0105} \\
\bottomrule
\end{tabular}
\end{table}

\begin{figure}[htb!]
    \centering
    \subfloat[]{\includegraphics[width=0.32\textwidth]{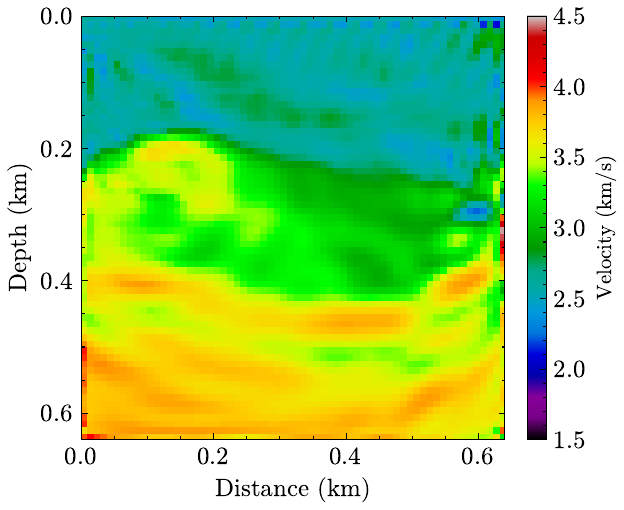}}\label{fig:left_shot_conv}
    \subfloat[]{\includegraphics[width=0.32\textwidth]{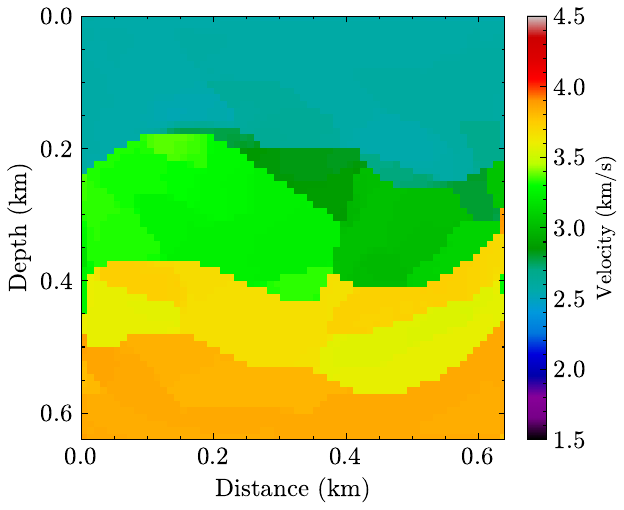}}\label{fig:left_shot_diffusion}
    \subfloat[]{\includegraphics[width=0.32\textwidth]{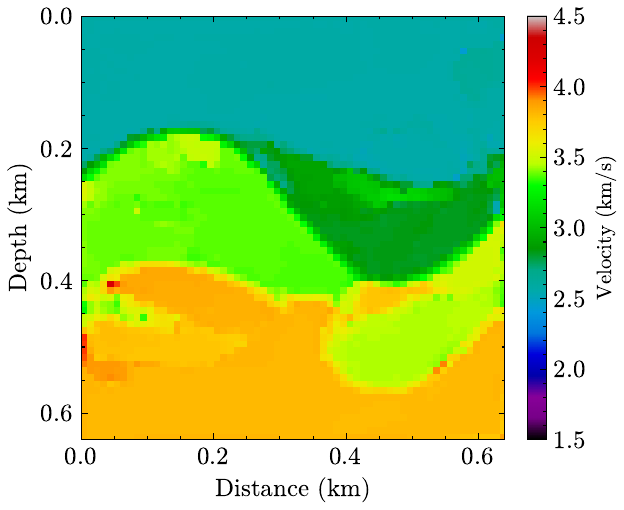}}\label{fig:left_shot_diffusion_class}
    \subfloat[]{\includegraphics[width=0.32\textwidth]{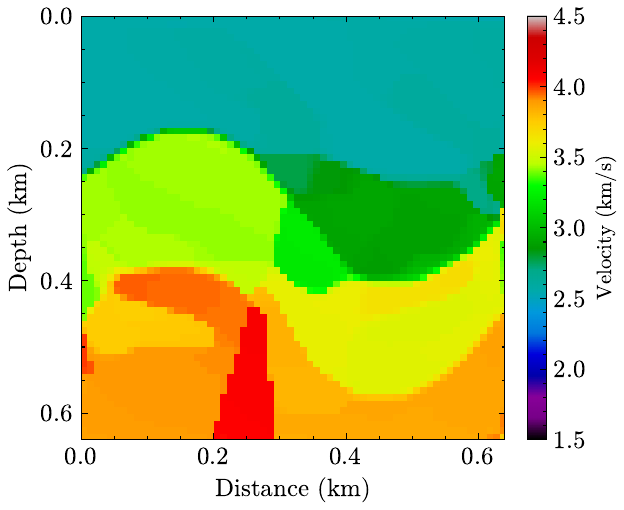}}\label{fig:left_shot_diffusion_well}
    \subfloat[]{\includegraphics[width=0.32\textwidth]{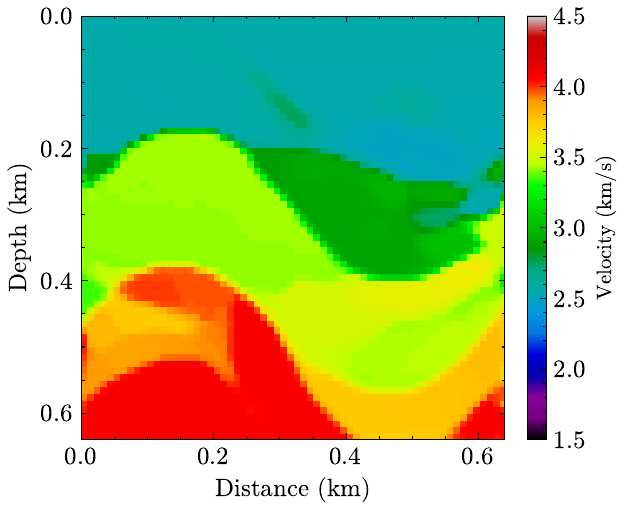}}\label{fig:left_shot_diffusion_well_class}
    \caption{Inversion results with a limited and closely-spaced shot array from a) conventional FWI, b) diffusion FWI, and conditional diffusion FWI with c) the geological class label, d) a well located at 0.25 km, and e) the geological class label and the well information together. Under limited observation conditions, both conventional FWI and standard diffusion-based FWI fail to produce accurate results, whereas the proposed method achieves significantly better performance. The diffusion FWI, conditioned on both well and class information, delivers the most accurate inversion.}
    \label{fig:left_shot}
\end{figure}

In a more challenging scenario, we applied conditional diffusion FWI with a limited and closely-spaced shot array: four shots are placed at 0 km, 0.02 km, 0.04 km, and 0.06 km, which correspond to the far left side of the model, which should cause poor illumination of the right side of the model. The inversion results are presented in Figure \ref{fig:left_shot}. 
As a result, the conventional FWI fails to recover the subsurface velocity, especially on the right side. The same holds for diffusion FWI, but with generally higher resolution. Incorporating the geological class label, as shown in Figure \ref{fig:left_shot}c, improved the subsurface structures. On the other hand, adding well-log information (Figure \ref{fig:left_shot}d) enhanced the velocity estimation at depth. The well information helps guide the velocity distribution with depth, especially near the well location. Meanwhile, the geological class information has a more global effect on the prediction, especially on the structure (less on the values). Using the geological class and well condition simultaneously, we obtain the inversion result in Figure \ref{fig:left_shot}e, which is the closest to the true velocity among the five results.
The quantitative statistics in Table~\ref{tab:model_comparison_limited_data} further demonstrate that the proposed approach consistently outperforms conventional FWI and diffusion-based FWI methods.
\begin{table}[h]
\centering
\caption{Quantitative comparison of different approaches with a limited and closely-spaced shot array. Higher PSNR and SSIM, lower relative L2 error are better.}
\label{tab:model_comparison_limited_data}
\begin{tabular}{lccc}
\toprule
\textbf{Approaches} & \textbf{PSNR (dB)}$\uparrow$  & \textbf{SSIM}$\uparrow$ & \textbf{Relative L2 error}$\downarrow$ \\
\midrule
\texttt{Conventional FWI}                   & 25.31 & 0.468 & 0.0736 \\
\texttt{Diffusion FWI}                  & 25.47 & 0.543 & 0.0630 \\
\texttt{CD-FWI (class)}       & 26.83 & \textbf{0.649} & 0.0586 \\
\texttt{CD-FWI (well)}        & 27.80 & 0.623 & 0.0504 \\
\texttt{CD-FWI (class\&well)} & \textbf{28.15} & 0.623 & \textbf{0.0485} \\
\bottomrule
\end{tabular}
\end{table}
\begin{figure}[htb!]
    \centering
    \subfloat[]{\includegraphics[width=0.32\textwidth]{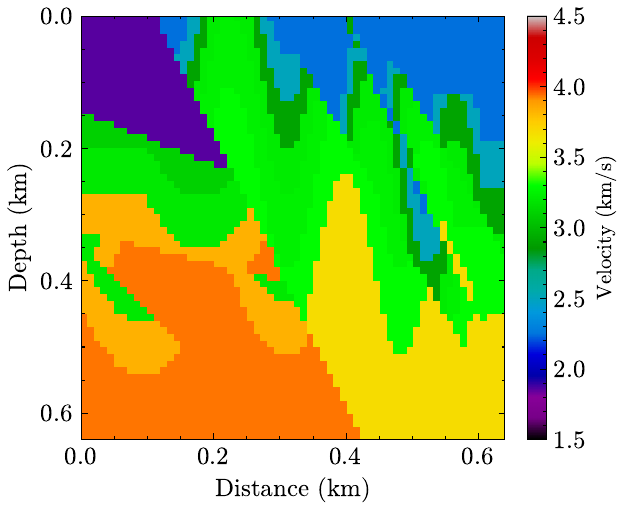}}\label{fig:well_model_true}
    \subfloat[]{\includegraphics[width=0.32\textwidth]{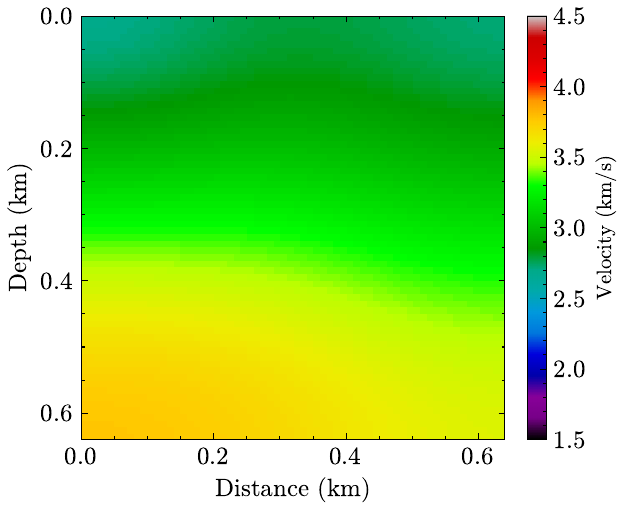}}\label{fig:well_model_init}
    \caption{a) The true velocity model, and b) the initial velocity model.}
    \label{fig:well_true&init}
\end{figure}

\begin{figure}[htb!]
    \centering
    \subfloat[]{\includegraphics[width=0.32\textwidth]{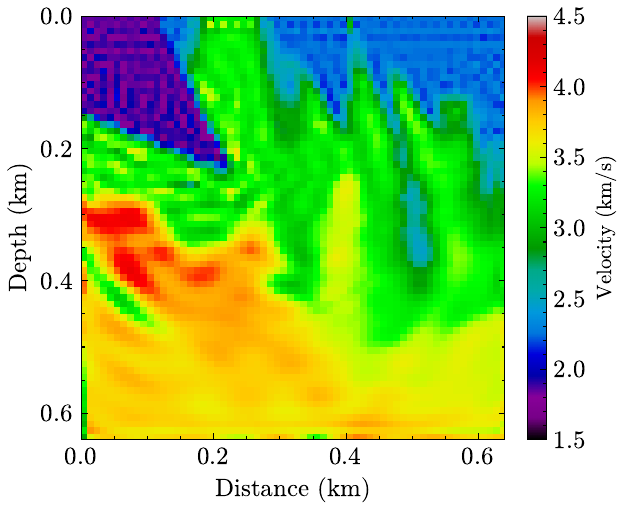}}\label{fig:well_conv}
    \subfloat[]{\includegraphics[width=0.32\textwidth]{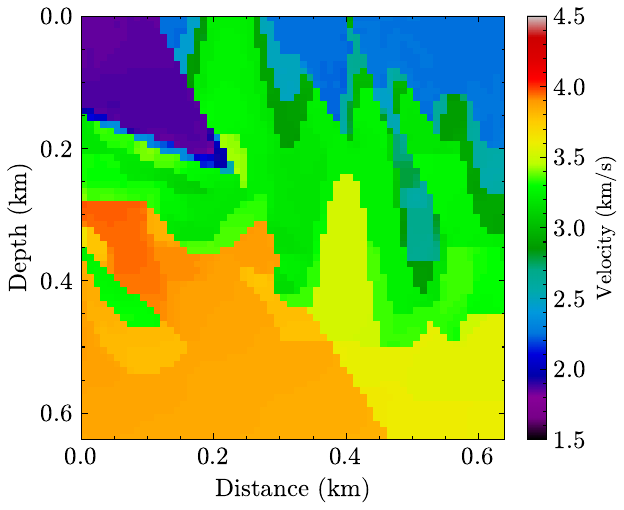}}\label{fig:well_diffusion}\\
    \subfloat[]{\includegraphics[width=0.32\textwidth]{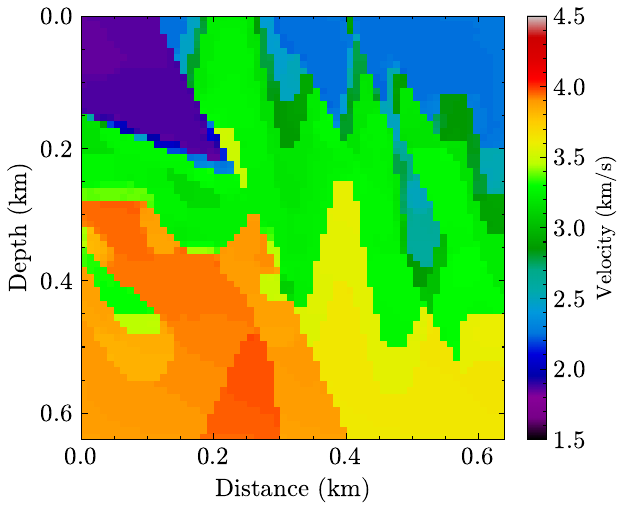}}\label{fig:well_diffusion_25}
    \subfloat[]{\includegraphics[width=0.32\textwidth]{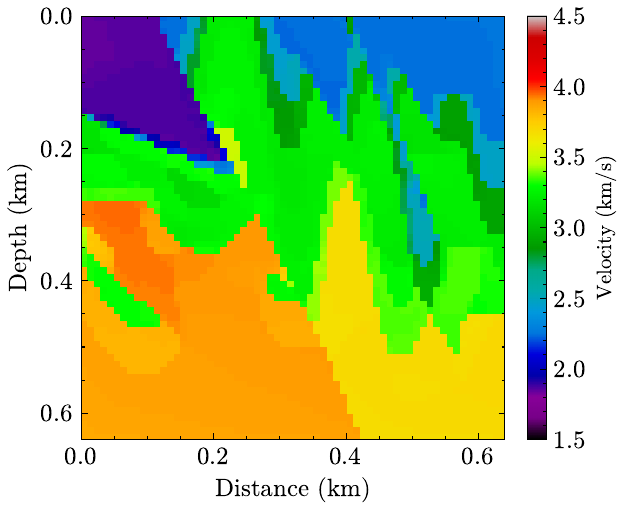}}\label{fig:well_diffusion_well_40}
    \caption{Inversion results from a) conventional FWI, b) diffusion FWI, and well-conditioned diffusion FWI (c), and (d) with the well located at 0.25km and 0.4km, respectively.}
    \label{fig:well_test}
\end{figure}

\begin{figure}[htb!]
    \centering
    \subfloat[]{\includegraphics[width=0.75\textwidth]{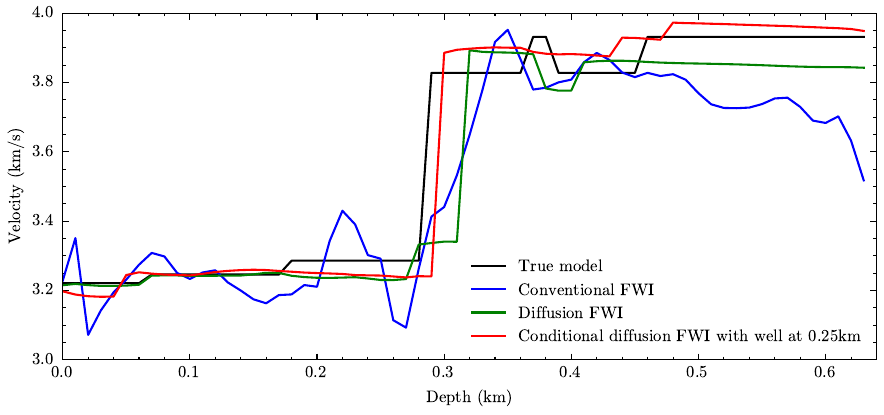}}\label{fig:well_profile_4_25}
    \subfloat[]{\includegraphics[width=0.75\textwidth]{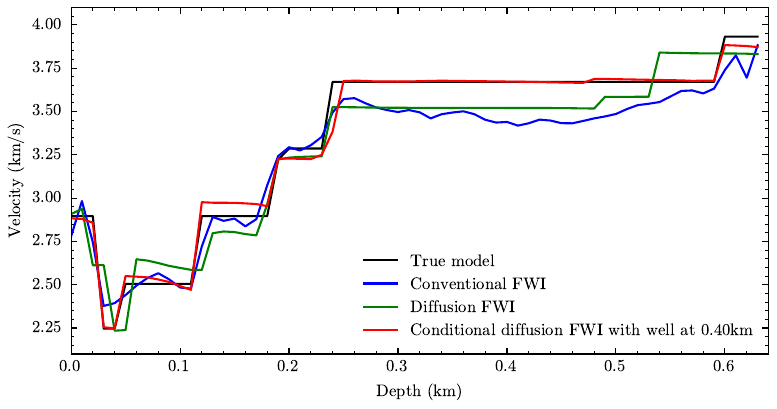}}\label{fig:well_profile_4_40}
    \caption{Profile comparison for the three well-conditioned diffusion FWI results (a) at 0.25km (Figure \ref{fig:well_test}c),  
    and (b) at 0.40km (Figure \ref{fig:well_test}d) with the true model in Figure \ref{fig:well_true&init}a, conventional FWI results in Figure \ref{fig:well_test}a, and diffusion FWI result in Figure \ref{fig:well_test}b, respectively. It highlights the importance of the well location for the accuracy of the well-conditioned diffusion FWI.}
    \label{fig:well_condition_comparison}
\end{figure}

From the above tests, we observe clear improvements with geological and well log priors-assisted FWI. 
Unlike the class label, the well log depends on an important factor, which is the location of the well log. 
To further test and understand the influence of the well location on the diffusion FWI, we select another test (not used in the training) velocity model shown in Figure \ref{fig:well_true&init}. 
This velocity model includes a very complex fault structure and strong lateral velocity variations from shallow to deep. 
It will help us test the well-log information guidance for complex scenarios.
The experiment parameters regarding the geometry and source configurations are the same as the test in Figure \ref{fig:middle_shot}. 
We conduct conventional FWI, diffusion FWI, and well-conditioned diffusion FWI with the well logs at different locations, whose results are shown in Figure~\ref{fig:well_test}, and the corresponding quantitative comparison is shown in Table~\ref{tab:model_comparison_well}. 
The corresponding profile comparisons for well condition at each position are shown in Figure \ref{fig:well_condition_comparison}. 
We noticed that the well-log information can improve the velocity values, especially at depth. However, due to the strong velocity changes in the lateral direction, this improvement is mainly tied to the well location.
An improper well-log location will somehow limit its guidance in diffusion-regularized FWI, resulting in relatively poor inversion results.
So, this experiment again demonstrates that the best way to condition diffusion FWI in a complex scenario is to include local and global type conditions, specifically well logs and class labels, in our experiment.
This makes sense as the more information considered, the better results we should obtain, which is the main motivation of this paper. 
\begin{table}[h]
\centering
\caption{Quantitative comparison of different approaches with different conditions (shown in Figure~\ref{fig:well_test}). Higher PSNR and SSIM, lower relative L2 error are better.}
\label{tab:model_comparison_well}
\begin{tabular}{lccc}
\toprule
\textbf{Approaches} & \textbf{PSNR (dB)}$\uparrow$  & \textbf{SSIM}$\uparrow$ & \textbf{Relative L2 error}$\downarrow$ \\
\midrule
\texttt{Conventional FWI}                   & 27.51 & 0.727 & 0.0528 \\
\texttt{Diffusion FWI}                  & 29.21 & 0.786 & 0.0417 \\
\texttt{CD-FWI (well at 0.25 km)}        & 29.57 & 0.788 & 0.0401 \\
\texttt{CD-FWI (well at 0.40 km)} & \textbf{29.96} & \textbf{0.810} & \textbf{0.0380} \\
\bottomrule
\end{tabular}
\end{table}

\section{Field data application} 
\label{field}
In this section, we applied our method to field data to further validate the effectiveness of the geological and well-prior conditioned diffusion regularized FWI.
We specifically test the proposed method on 2D marine streamer data acquired from the North-Western Australian Continental Shelf by CGG. 
To reduce the computational cost of FWI, we select 116 shots with a lateral interval of 98.75 $m$ from the original 1824 shots with an interval of 18.75 $m$.
The receiver gather includes 648 receivers with a spacing of 12.5 $m$, yielding a minimum and maximum offset of 16.9 $m$ and 8256 $m$, respectively. 
The data were recorded for a maximum duration of about 7 seconds with a 2-$ms$ sampling interval. 
The velocity model covers an area of 12.5 $km$ in length and 3.7 $km$ in depth, with a grid interval of 12.5 $m$ in both $x$ and $z$ directions. Similar to our previous preprocessing \citep{wang2023prior}, we first applied a low-pass filter on the field data to filter out the component above 8.0 Hz. 
The source wavelet is inverted from the near-offset early arrival seismograms with a constant water velocity of 1500 m/s. Then, the following sequences of processing steps, including a brute-stack time-domain velocity analysis \citep{kalita2019flux} followed by an extended full waveform inversion with matching filter \citep{li2021extended}, were used to obtain a good initial model (shown in Figure~\ref{fig:inversion_test}a).
To highlight the influence of priors with different modalities on the inverted results using our method, we only test the single prior conditioned diffusion FWI in this section.

\begin{figure}[htb!]
    \centering
    \includegraphics[width=0.75\textwidth]{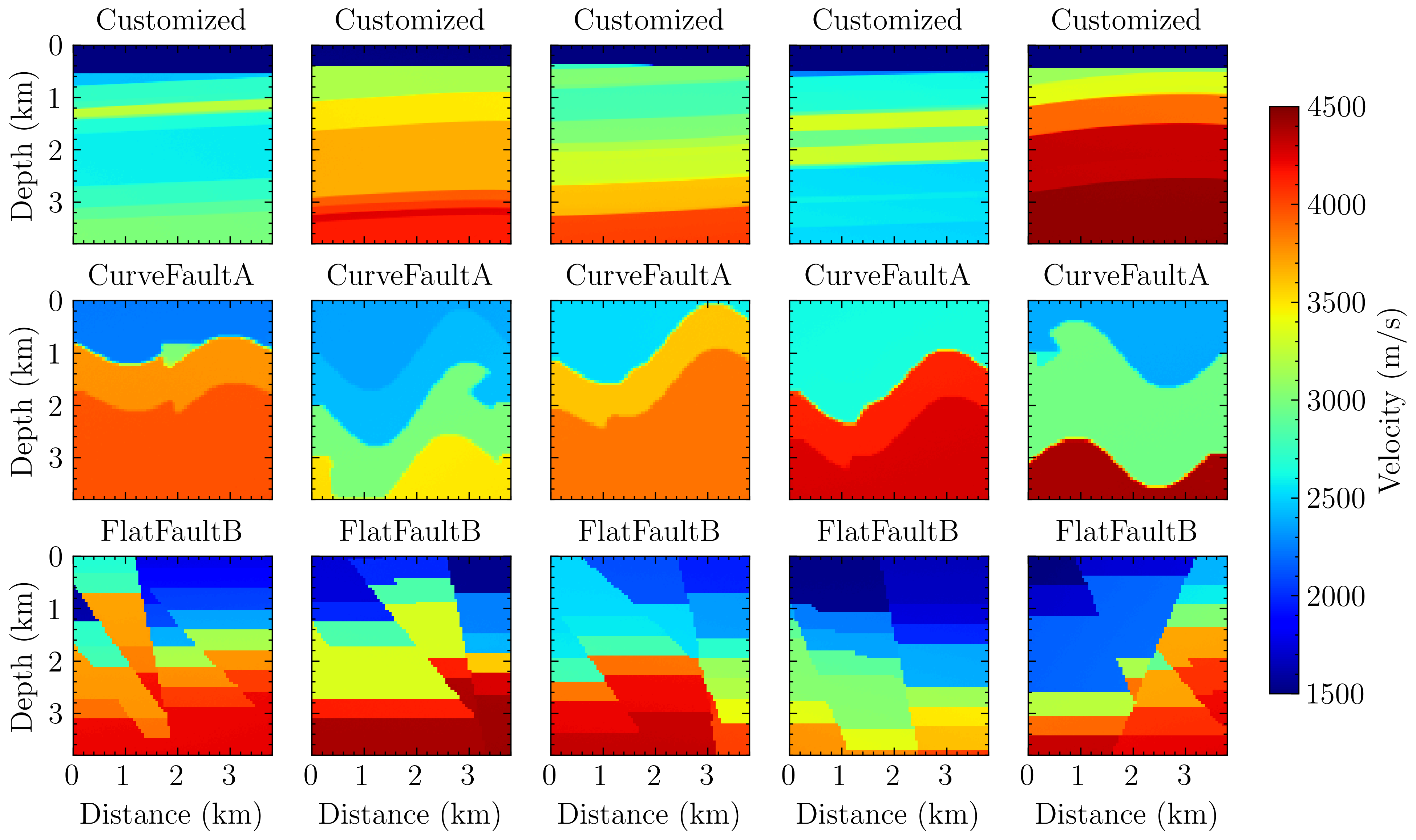}
    \caption{The samples of generated velocity models using conditional diffusion models given the three class conditions prepared for the field data.}
    \label{fig:class_velocity}
\end{figure}
\begin{figure}[htb!]
    \centering
    \subfloat[]{\includegraphics[width=0.68\textwidth]{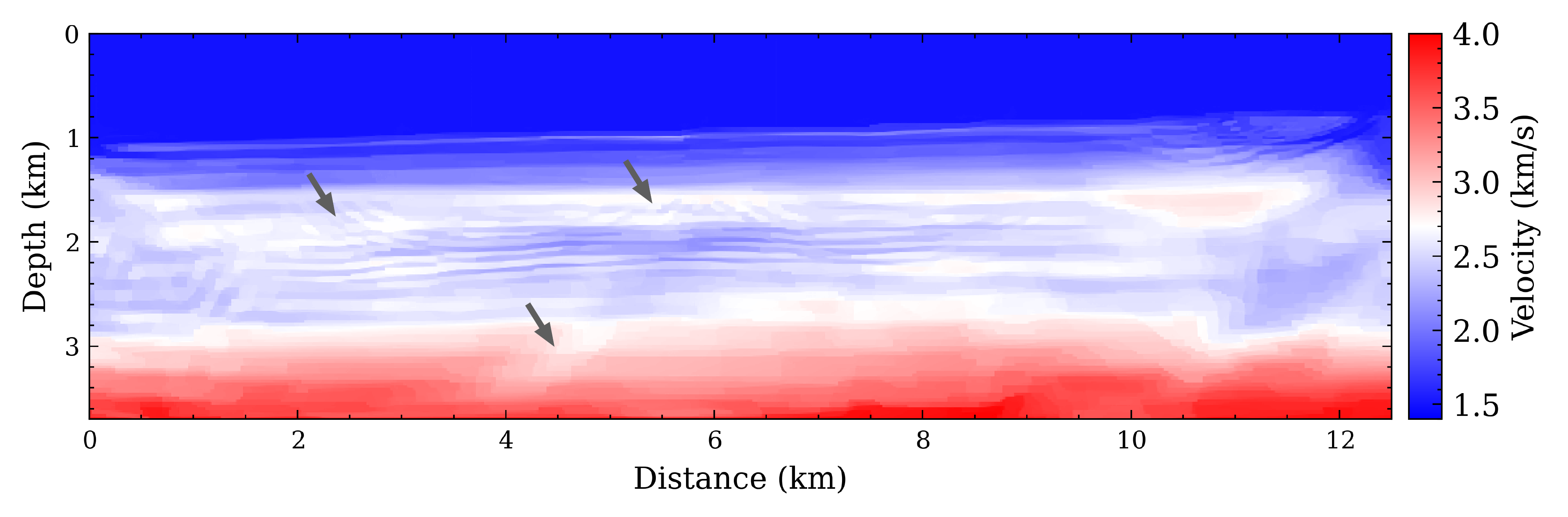}}\
    \subfloat[]{\includegraphics[width=0.68\textwidth]{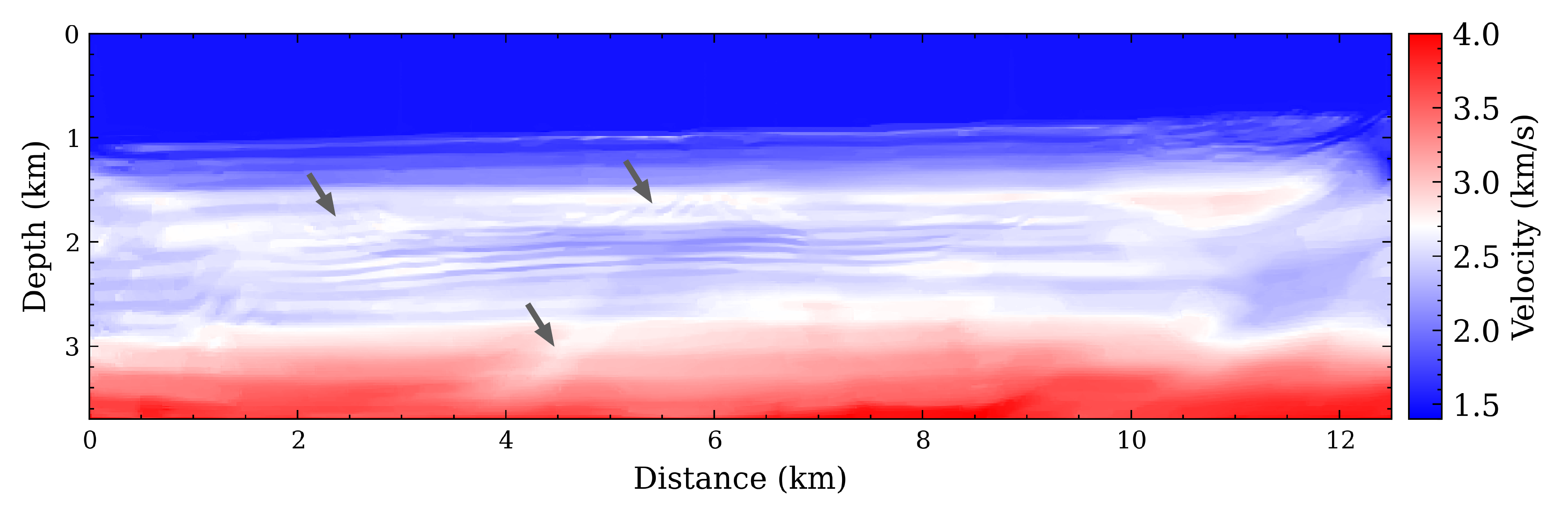}}\
    \subfloat[]{\includegraphics[width=0.68\textwidth]{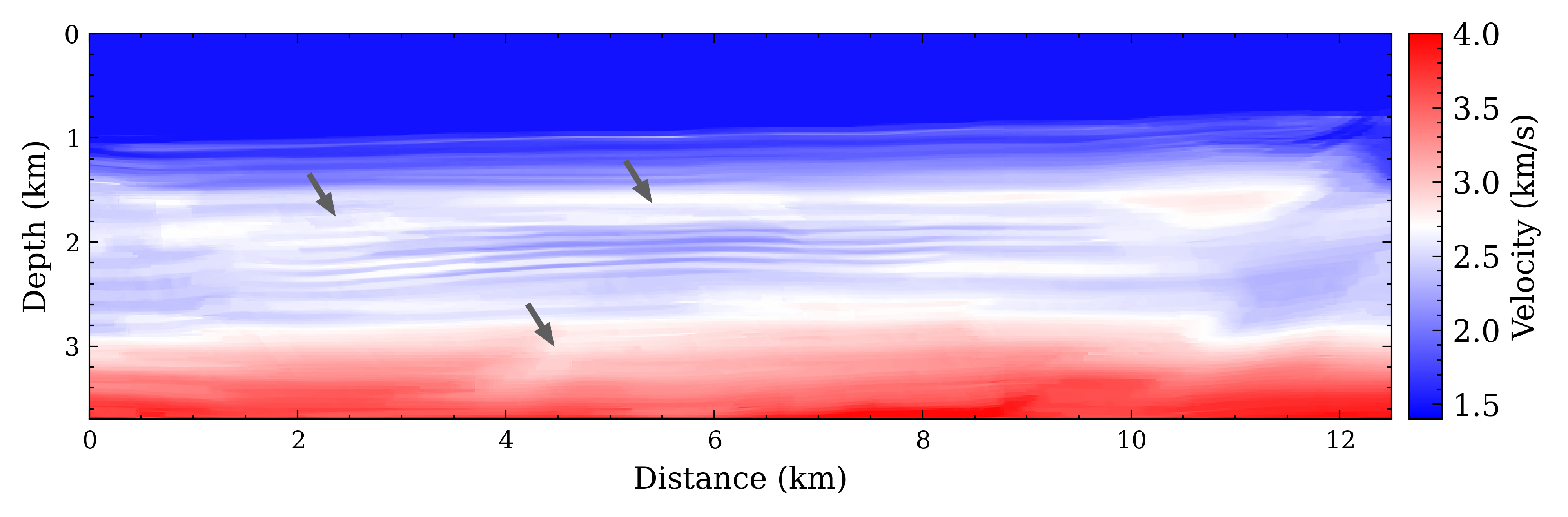}}
    \caption{The inverted results using the patch-based class conditioned diffusion regularized FWI. The result with a class condition of "FlatFaultB" (a), that with "CurveFaultA" (b), and that with "Customized" (c). The arrows point to areas of differences in which (a) and (b) show more fault features and artifacts. These results demonstrate that conditional diffusion FWI improves inversion accuracy and that incorporating suitable prior information as conditioning leads to further enhancements.}
    \label{fig:class_field}
\end{figure}

\subsection{Geological prior assisted FWI} 
We first test the geological prior assisted FWI on the field data.
Since we do not have an exact geological description of the field subsurface, we created the datasets of velocity models with relatively flat or mild curvature features inside the model and with a resolution of $152\times152$, marked as a class "Customized". 
Then, we combine this class with two classes from the OpenFWI dataset: "CurveFaultA" and "FlatFaultB" (for which we upsample the resolution to $152\times152$) to have a total of three classes in our training data. 
The U-Net model has demonstrated strong capacity across different OpenFWI datasets, which encompass a variety of geological complexities such as curvatures and fault structures. 
For the field data case, we adopt a patch-based approach, and we found that the current U-Net architecture has sufficient capacity to effectively handle localized structures within each patch. 
Hence, we train a conditional diffusion model using these three classes with the same training configuration mentioned in the synthetic tests.
Because the training of pixel space diffusion model on high-resolution images, like the model size considered for this FWI, is hard and costly, instead, we train the diffusion model using a resolution of 152$\times$152 and then use patch-based strategies during the FWI \citep{wang2023prior}.
The sampled results using the trained diffusion model conditioned on three classes are shown in Figure~\ref{fig:class_velocity}.

Here, we use 11 reverse diffusion steps and do 8 iterations of FWI within each diffusion step. 
The step length is 0.002 for the FWI update.
As the size of the velocity model is 296$\times$1000 grid points, directly inputting the velocity model into the diffusion model originally trained on the resolution of 152$\times$152 will result in artifacts.
Instead, we use a patch-based strategy to extract six horizontal patches of size 296$\times$296 and then merge them back to the original velocity model size.
The results using different geological priors are shown in Figure~\ref{fig:class_field}. 
Although the initial velocity model has already provided sufficient information on the background to FWI, different geological priors will result in different inverted results.
The class label "FlatFaultB" promotes more flatness and lateral discontinuity (Figure~\ref{fig:class_field}a) and turns the curved areas into faults.
Using the class label "CurveVelA" as a condition, more curvatures (Figure~\ref{fig:class_field}b) appear in the inverted result, but the artifacts in the horizon direction are hard to suppress.
When using the class label "Customized", the higher wavenumber components of the subsurface model are reconstructed with better horizontal continuity, as shown in Figure~\ref{fig:class_field}c. 
With this better geological prior, the conditional diffusion regularized FWI achieves the best result. 
Further waveform misfit analysis (Figure~\ref{fig:field_convergence_curve}) will demonstrate this observation as well.

This experiment demonstrates that if we could have better geological priors that can exactly describe the subsurface structures, e.g., the information on whether the subsurface tends to be more curved or flat or whether there is a fault, we can obtain a more reasonable inverted result. 
In addition, this field application further demonstrates the feasibility of the proposed method to let the geologist join the FWI process and the importance of incorporating geological information.

\subsection{Well conditioned diffusion regularized FWI}
\begin{figure}[htb!]
    \centering
    \includegraphics[width=0.51\textwidth]{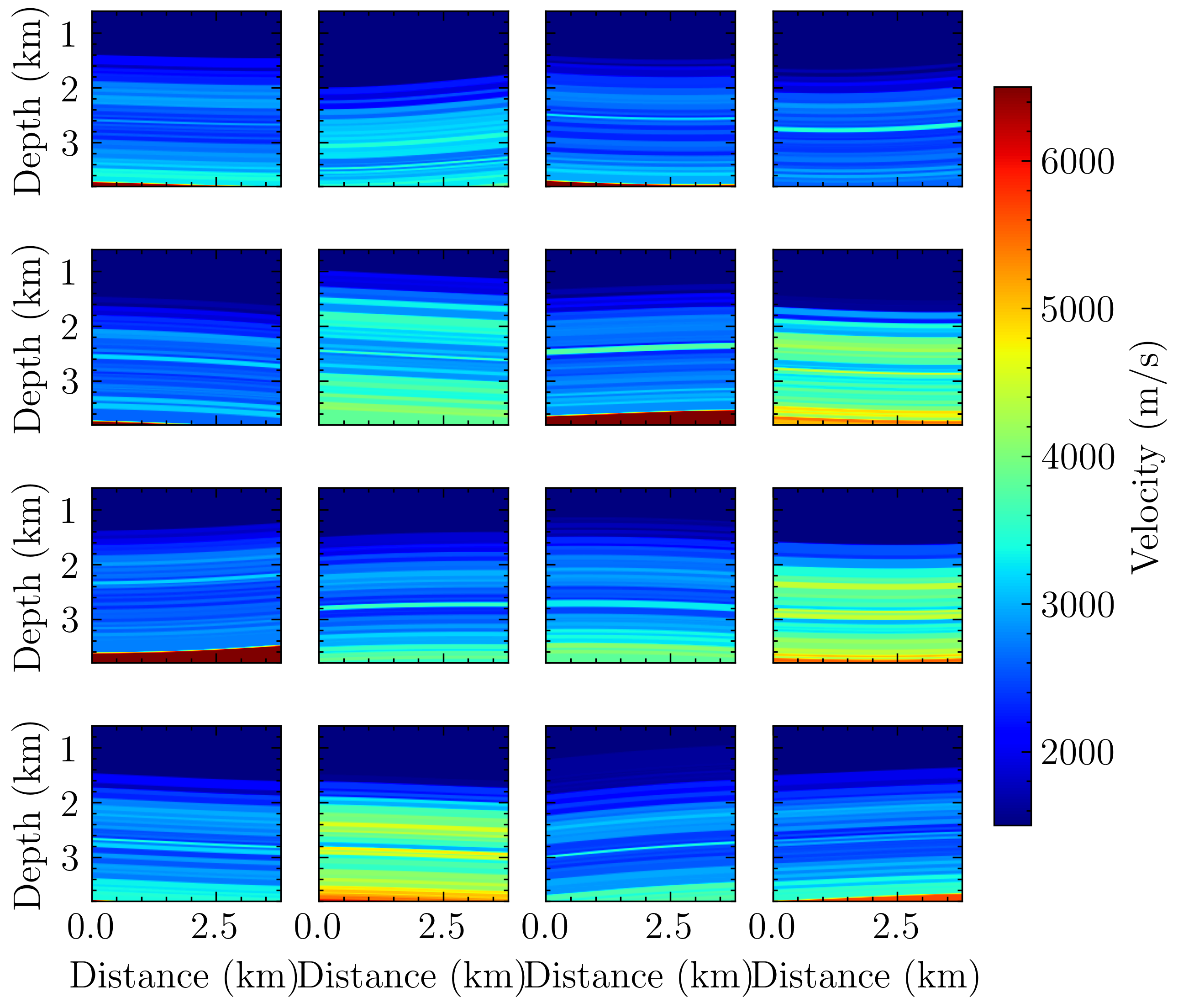}
    \caption{The samples of the generated high-resolution velocity models guided by the well log and prior knowledge of the subsurface structures. We use these velocity models to train the conditional diffusion models by selecting random vertical velocity profiles as conditions.}
    \label{fig:high_resolution velocity}
\end{figure}

\begin{figure}[htb!]
    \centering
    \subfloat[]{\includegraphics[width=0.51\textwidth]{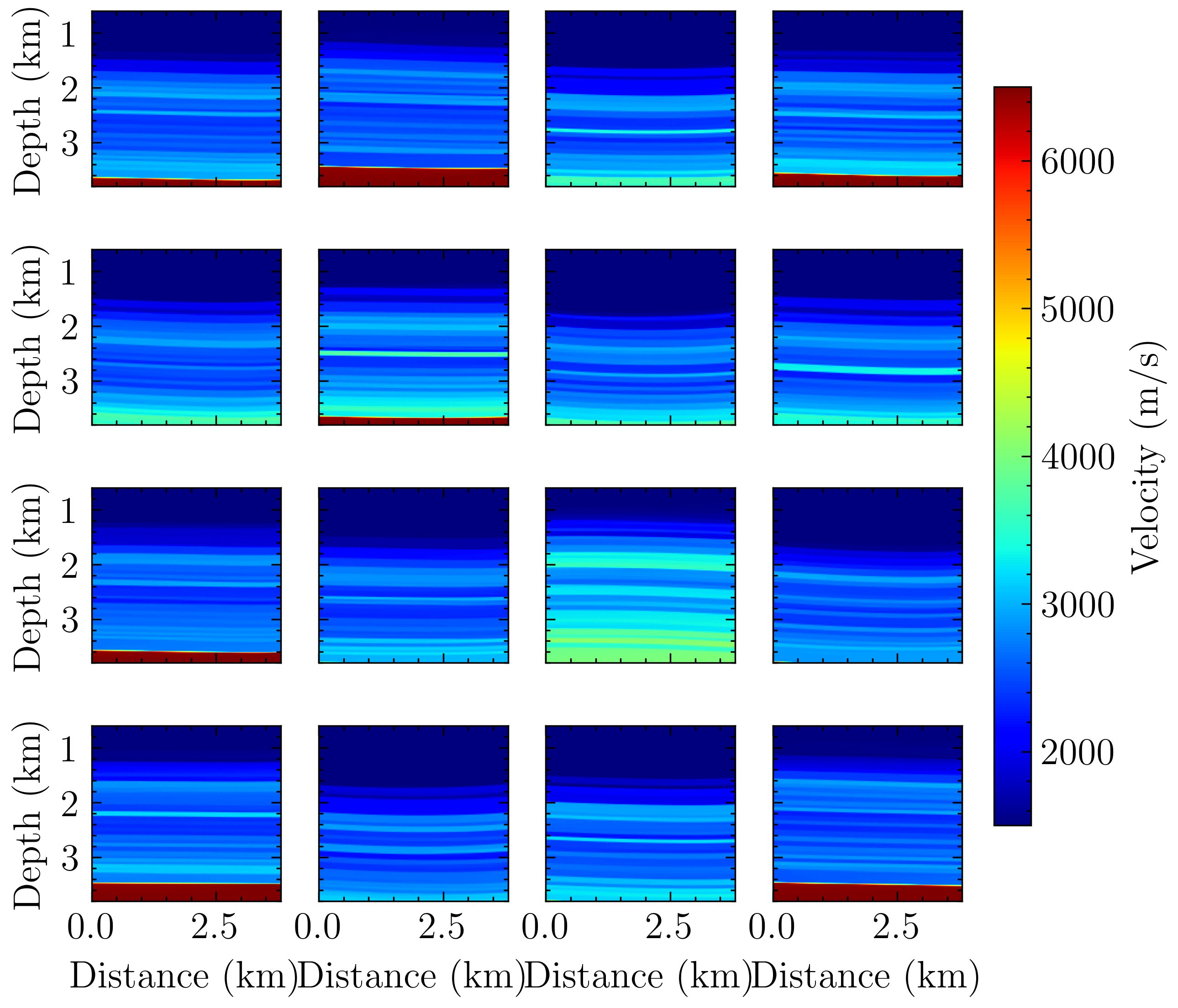}}\
    \subfloat[]{\includegraphics[width=0.42\textwidth]{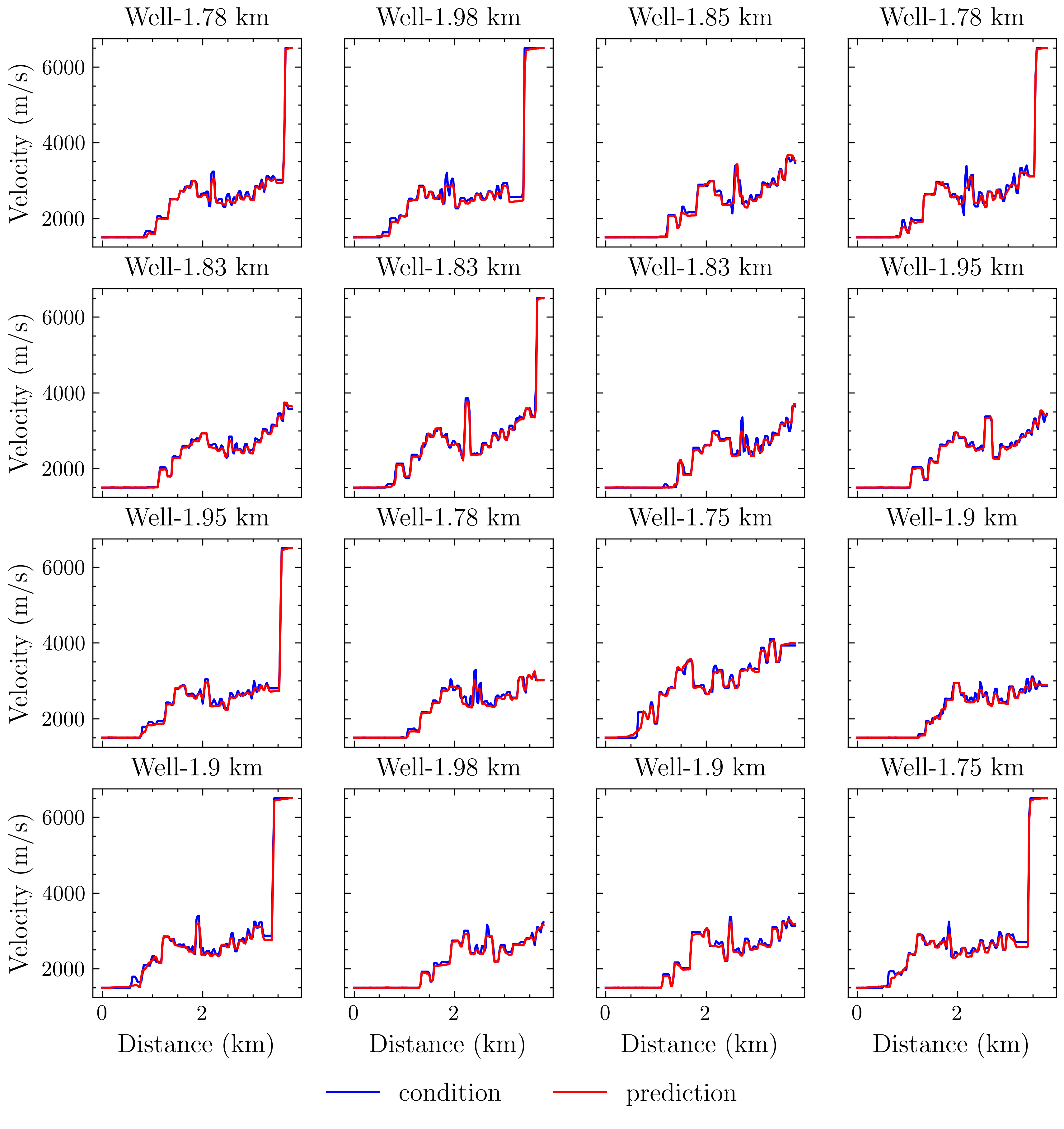}}    
    \caption{(a) 16 samples of the generated velocity using conditional diffusion model, and (b) the corresponding comparison between the extracted well profile from these generated velocity models and given conditions.}
    \label{fig:well_training}
\end{figure}

In this subsection, we test the well-log conditioned diffusion regularized FWI.
We generate 80000 samples with a resolution of $152\times152$ guided by the vertical profiles and structural prior \citep{ovcharenko2022multi}.
The samples of the generated velocity models for training are shown in Figure~\ref{fig:high_resolution velocity}.
We use the same configuration as the OpenFWI test earlier to train the conditional diffusion model. 
After training the conditional diffusion model, we can sample the velocity models given well-log (velocity profile) conditions. 
Figure~\ref{fig:well_training} shows 16 samples generated from the conditional diffusion model given different well-log profiles.
We observe that the matching between the well-log profiles and profiles extracted from the predicted velocity models is generally good.

\begin{figure}[htb!]
    \centering
    \subfloat[]{\includegraphics[width=0.68\textwidth]{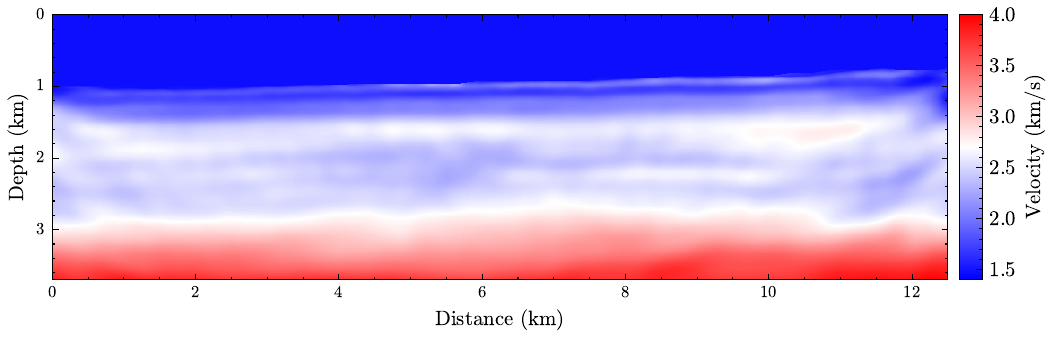}}\label{fig:init}
    \subfloat[]{\includegraphics[width=0.68\textwidth]{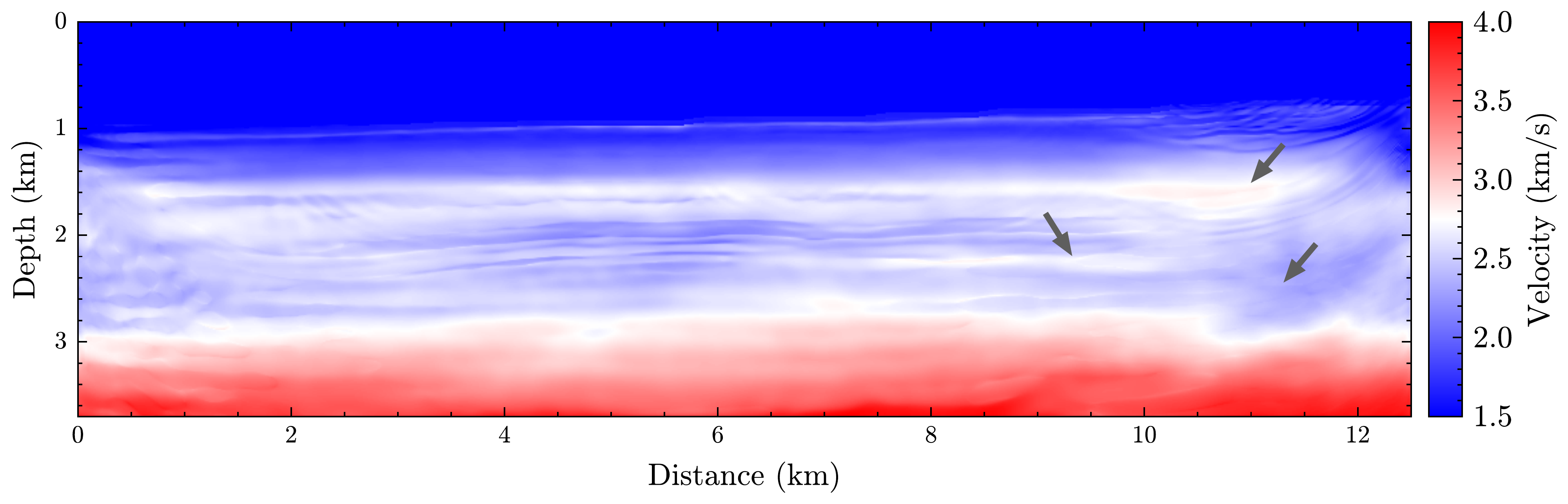}}\label{fig:convFWI}
    \subfloat[]{\includegraphics[width=0.68\textwidth]{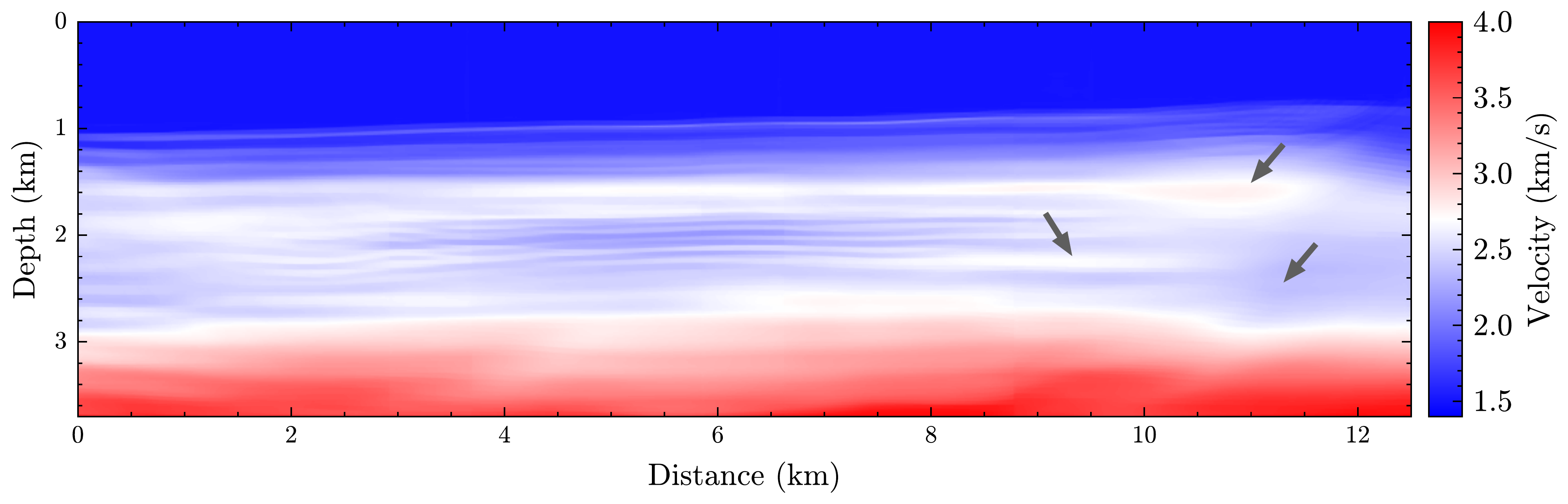}}\label{fig:diffusionFWI}
    \subfloat[]{\includegraphics[width=0.68\textwidth]{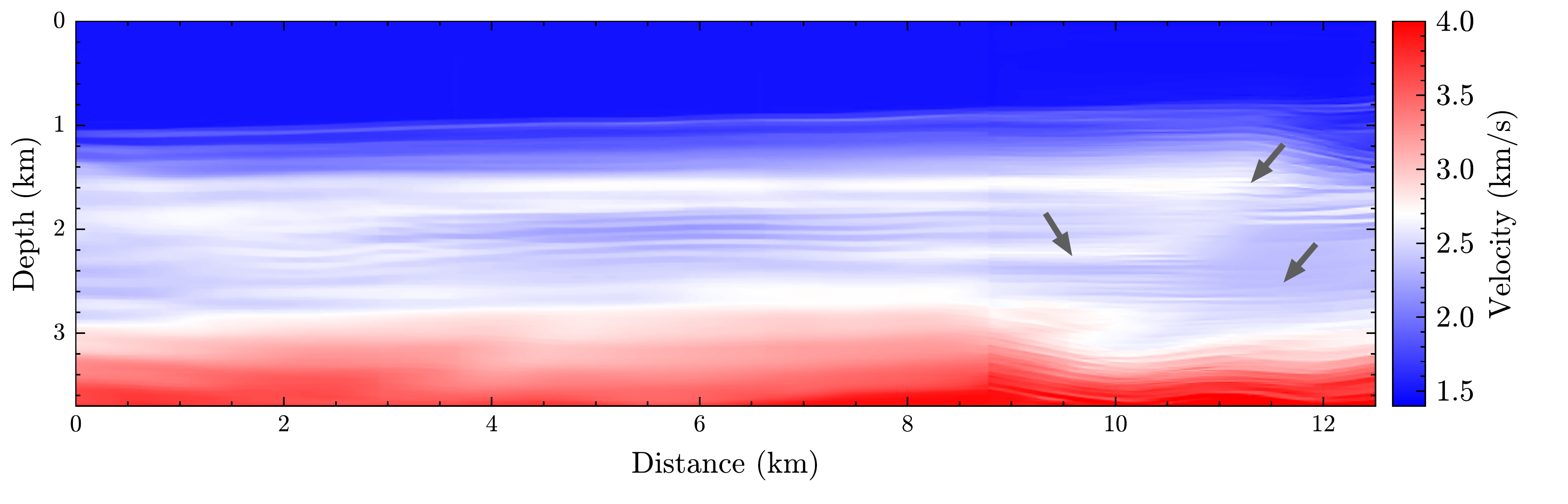}}\label{fig:cdiffusionFWI}   
    \caption{Starting from the initial velocity model (a), we obtain the inversion results from (b) conventional FWI,  (c) diffusion FWI where the conditional factor for well log is -1 (no well log information), and (d) well-conditioned diffusion FWI where the conditional factor for well log is 19. Compared with the inversion result in (b) and (c), the well-conditioned diffusion FWI can admit higher wavenumber components in (d), especially in the areas denoted by the arrows. The well-log guidance can somehow adjust the velocity value closer to the well log and match the fluctuations in the velocity profile.}
    \label{fig:inversion_test}
\end{figure}

\begin{figure}[htb!]
    \centering
    \includegraphics[width=0.75\textwidth]{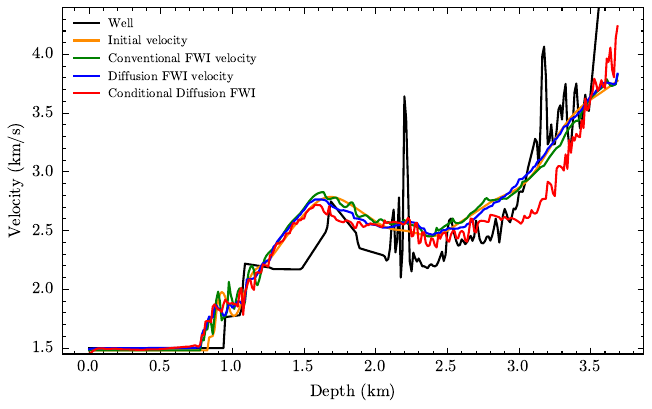}
    \caption{The vertical profiles comparison at the well location of 10.5$km$, of the initial velocity model, and the inverted results using the conventional FWI, diffusion regularized FWI, and conditional diffusion regularized FWI. }
    \label{fig:well_comp}
\end{figure}

After we have tested the trained diffusion model to generate velocities consistent with the trained distribution and close to the provided well information, we use this model in FWI. 
Similar to the geological prior conditioned diffusion regularized FWI test, we split the subsurface velocity model into 6 patches and give only the patches that cover the well log location with the condition scale $\lambda$ equal to 19 (denoting conditional generation), while others with -1 (denoting the unconditional generation). 
For unconditional diffusion FWI, the condition scale is equal to -1 for all patches.
As shown in Figure~\ref{fig:inversion_test}, since the initial model is the result of an extended FWI using a matching filter, the conventional FWI is generally able to converge, providing mainly (or only) high-resolution information (Figure~\ref{fig:inversion_test}b).
In contrast, the diffusion-regularized FWI and the well-conditioned diffusion-regularized FWI enhance the lateral continuity of the inverted model and remove the illumination artifacts at the edge of the velocity model.
Particularly for the result using the well-conditioned FWI, it recovers the higher wavenumber components of the velocity model.
We extract the profiles at the well location at a lateral position of 10.5 $km$ for comparison, which is shown in Figure~\ref{fig:well_comp}.
In general, the profile of the inverted result using the proposed method is closer to the well log than the one obtained using the conventional FWI, as well as the diffusion regularized FWI. 
Especially at the depth range from 1.5 to 2.8 $km$, the result using our method has obvious improvements. 
Similar to the results shown for the synthetic tests, the well-log guidance can somehow adjust the velocity value closer to the well log and match the fluctuations in the velocity profile.
However, we do observe that in some deeper regions, at a depth of 3.0 km, the inverted velocity from the conditional diffusion FWI appears biased toward lower values, despite showing improved alignment in terms of velocity fluctuations.
This discrepancy arises because the training dataset contains relatively low-velocity values in this depth range, which introduces a bias in the learned prior distribution. 
Since diffusion models implicitly capture and sample from the data distribution they are trained on, this prior preference influences the inversion outcome, even when well-log information is provided as a condition.
This behavior highlights a key property of the diffusion model to store the distribution of the training velocity models, and such a distribution will guide the inversion as it represents the prior distribution.
In addition, we can observe the effectiveness of the well-prior conditioned regularized FWI, and it generally performs better than other methods.
However, it also reveals a limitation: when the training distribution is not sufficiently diverse or representative (e.g., lacking higher velocities at certain depths), the model may be biased toward the dominant patterns in the training set.

\subsection{A Misfit Curve Analysis for Field Data}
\label{sec:misfit}
\begin{figure}
    \centering
    \includegraphics[width=0.6\linewidth]{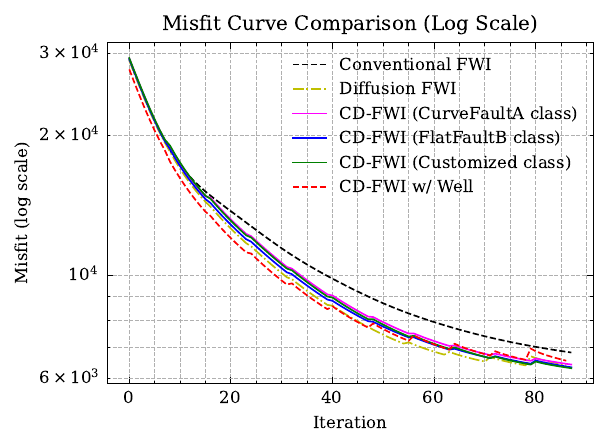}
    \caption{Misfit curves from field data inversions using different FWI strategies. The comparison includes conventional FWI, diffusion-based FWI, conditional diffusion FWI (CD-FWI) conditioned on different geological classes, and CD-FWI with well-log constraints. The vertical axis is plotted on a logarithmic scale to highlight convergence behavior across methods.}
    \label{fig:field_convergence_curve}
\end{figure}

To evaluate the effectiveness of different inversion strategies on real seismic data, we compare the misfit convergence curves obtained from a series of field data FWI experiments (see Figure~\ref{fig:field_convergence_curve}). The conventional FWI method exhibits slow but stable convergence, which is expected due to the high nonlinearity and lack of prior constraints. 
In contrast, the unconditional diffusion FWI accelerates convergence in early iterations by leveraging learned velocity priors from carefully constructed training set distributions, although it lacks control over geological structure.

The conditional diffusion FWI (CD-FWI) variants, conditioned on different geological class labels (e.g., CurveFaultA, FlatFaultB, and a customized class), demonstrate consistently improved convergence rates and lower final misfit values. 
This highlights the importance of incorporating geological context during the inversion process. 
Among them, CD-FWI with the customized class shows the best convergence performance, indicating that tailored conditioning aligned with the target data characteristics can further enhance inversion quality.

Moreover, the CD-FWI method incorporating well-log information achieves the lowest overall misfit at the earlier stages (within the first 50 iterations). 
This suggests that the integration of multi-modal prior constraints, particularly point-wise velocity constraints from well logs, provides significant benefits in stabilizing the inversion and guiding the optimization toward geologically plausible solutions.
In the later iterations, however, the waveform misfit does not reach the lowest values due to anisotropy in the region, leading to an inherent mismatch between the well-log velocity and the inverted FWI velocity.
If incorporating the anisotropic FWI or with better well-log data, it could be further improved.

\section{Discussion}
\label{discuss}
We used simple geological classes and well velocities to condition our diffusion-based generator prior. The results demonstrated the power of such conditioning in guiding the FWI results. There are no limits to including various information in the guidance, granted they are included in the training. Using more realistic models for training (industry-level models, not academia), we can include more high-resolution geological details like thin layers, salt bodies, and facies, and even specify their potential locations by incorporating localized classes. A form of such details can be accomplished by including facies maps along with the velocity models in training the diffusion models. For more structural guidance, we can include a migrated image obtained using the initial model as a condition. In summary, the potential for including prior information and interpreter input into FWI is limitless using diffusion models.

Regarding the additional cost, the proposed approach introduces minimal computational overhead during inference due to the additional diffusion regularized step. 
Specifically, each diffusion step takes approximately 0.036 seconds on an NVIDIA A100 GPU, which is negligible compared to the runtime of one iteration of conventional FWI. 
Although pretraining the diffusion model takes several hours, depending on the size of the training dataset and the number of iterations, it is performed entirely offline and only once.. 
As a result, the inference-time computational cost of our approach remains efficient and decoupled from the training stage. 
This separation enables practical deployment of the method without impacting the runtime efficiency of the inversion process.

\begin{figure}[htb!]
    \centering
    \subfloat[]{\includegraphics[width=0.32\textwidth]{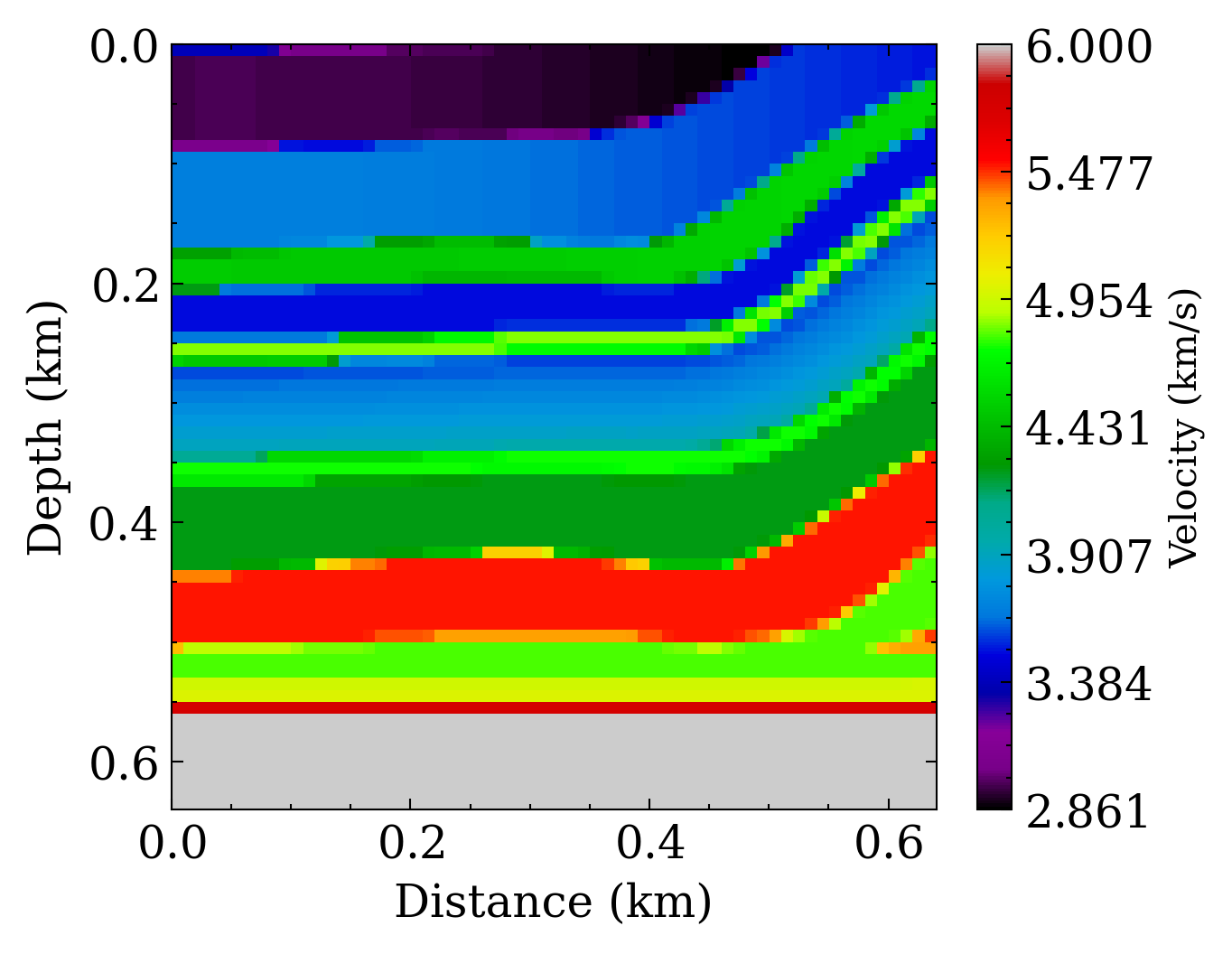}}\label{fig:model_true_overthrust}
    \subfloat[]{\includegraphics[width=0.32\textwidth]{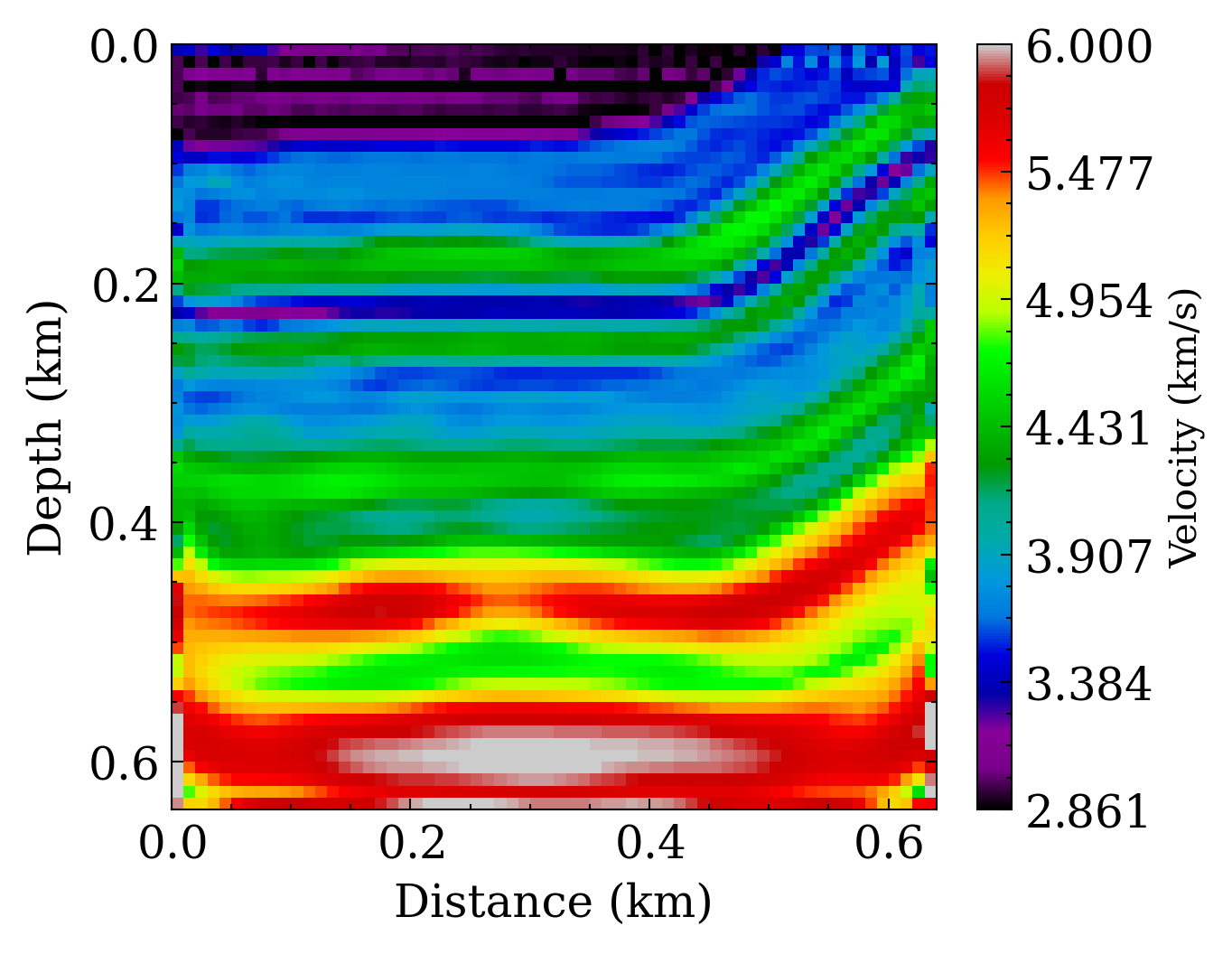}}\label{fig:model_class_overthrust}
    \subfloat[]{\includegraphics[width=0.32\textwidth]{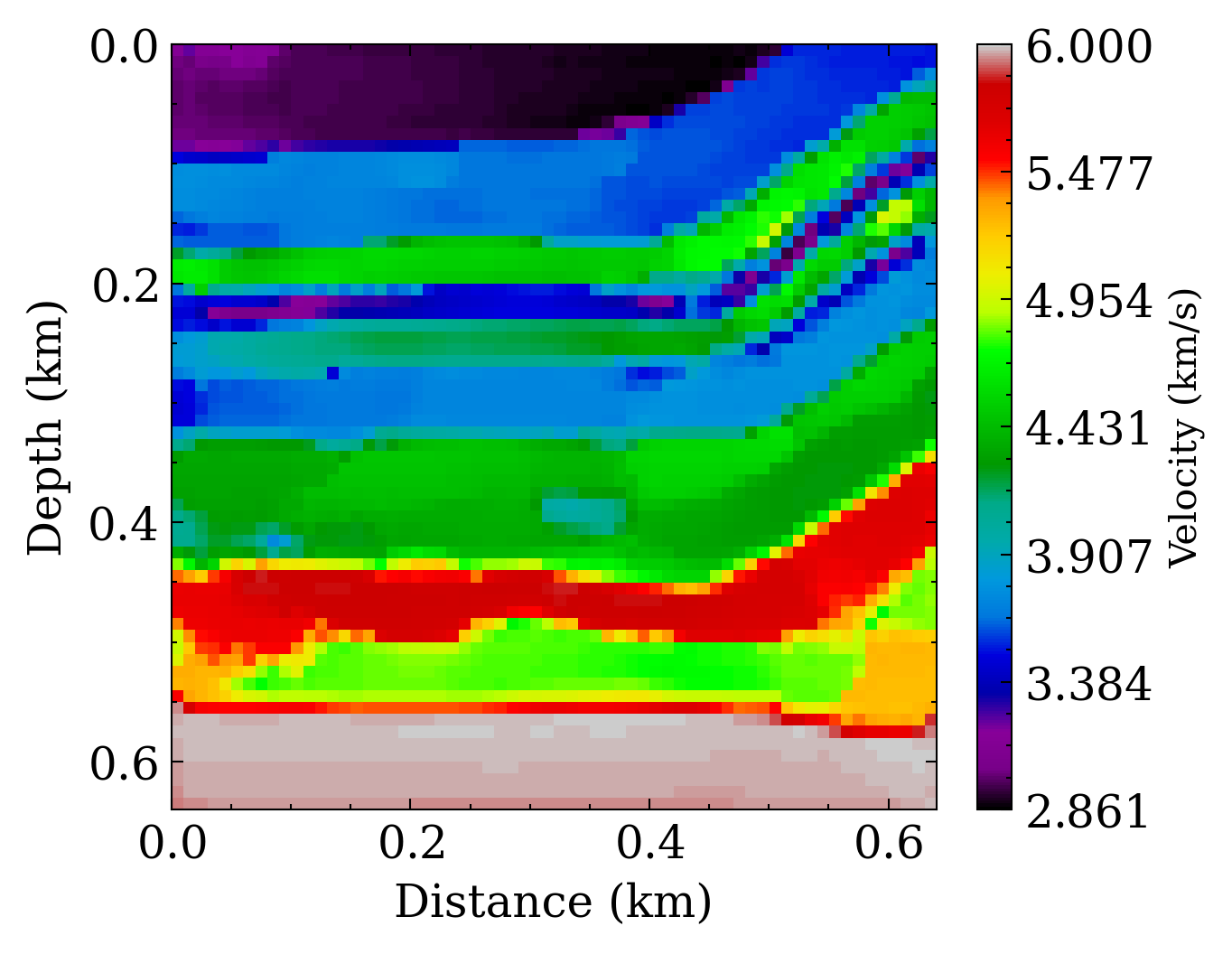}}\label{fig:model_well25_overthrust}
    \caption{ a)The true velocity model and inversion results from b) conventional FWI (whose PSNR, SSIM, and relative L2 error are 28.45dB, 0.78, and 0.05, respectively), c) conditional diffusion FWI with a well located at 0.25 km (whose PSNR, SSIM, and relative L2 error are 28.51dB, 0.80, and 0.04, respectively). The proposed conditional diffusion FWI generalizes well in such an out-of-distribution case.}
    \label{fig:overthrust}
\end{figure}

The proposed approach establishes a practical workflow: one can pre-train a diffusion model on a variety of generalized subsurface structures and subsequently apply it to field cases. The preceding field applications have already demonstrated the generalizability of the method, particularly in the context of class-conditioned diffusion FWI. Here, we further assess the generalizability of the well-conditioned diffusion FWI. Specifically, we apply the pre-trained diffusion model (trained on the OpenFWI dataset) to a portion of the Overthrust model (Figure~\ref{fig:overthrust}a), which differs significantly from the OpenFWI datasets in that it is not piecewise-constant and exhibits distinct geological features. The observation setup follows that of the CurveFaultB test, with the number of sources and receivers adjusted to 8 and 32, respectively. The initial model is derived by applying a Gaussian smoothing ($\sigma=20$) to the ground truth velocity. Figure~\ref{fig:overthrust} presents the inverted results from both conventional FWI and the proposed method. Despite being applied to an out-of-distribution scenario, the proposed approach exhibits strong performance, particularly in capturing deep structural details and accurate velocity magnitudes. This further highlights the robustness and generalizability of the proposed method.

One potential limitation of the proposed method is high-resolution FWI due to the inherent challenge of high-resolution image synthesis by diffusion models without any latent representation.
Although the patch-based method has shown good performance, for well-prior conditioned diffusion FWI, it is hard to determine the patch size and stride for the patches, which would further influence the regularization from well log conditions on velocities far away from the well log location.
As demonstrated in previous synthetic tests, given the presence of few sharp lateral velocity changes as shown in Figure~\ref{fig:well_true&init}, using the well-log condition should guide the inversion of the velocity model beyond just near the well location. 
Here, we retrain a well-log-conditioned diffusion model for high-resolution controllable velocity generation to demonstrate this challenge. 

\begin{figure}[htb!]
    \centering
    \includegraphics[width=1.0\textwidth]{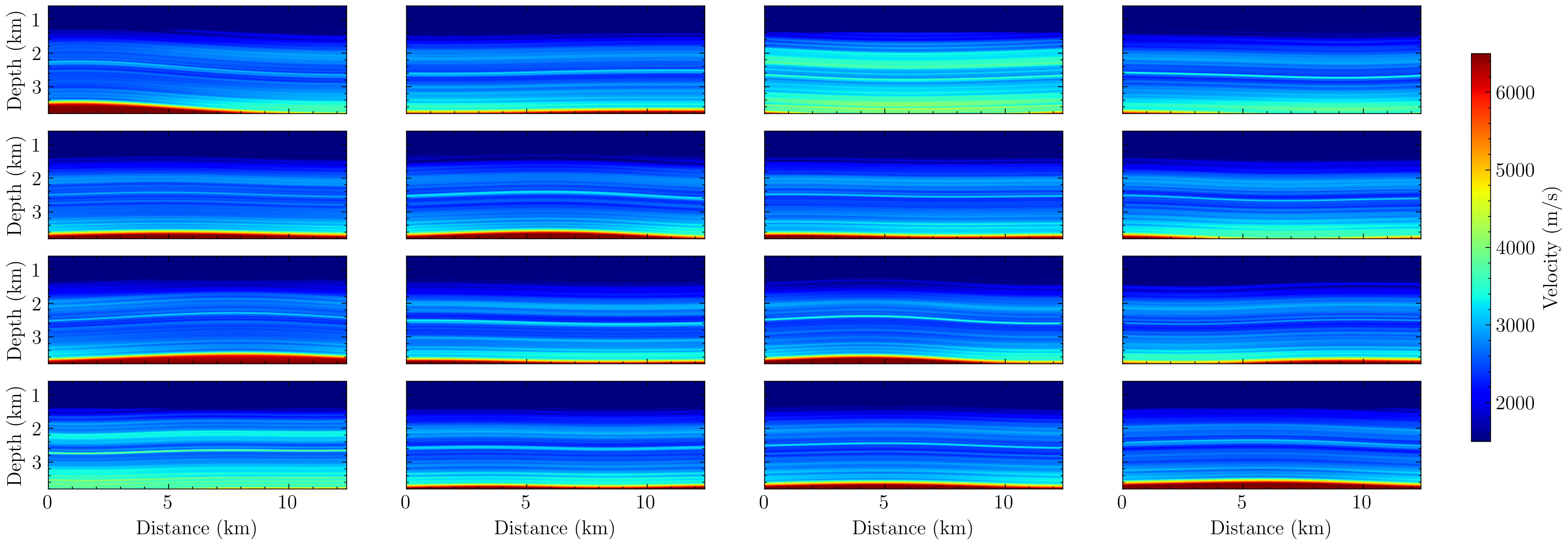}
    \caption{The samples of the generated high-resolution velocity models guided by the well log and prior knowledge of the subsurface structures. We use these velocity models to train the conditional diffusion models by selecting random vertical velocity profiles as conditions.}
    \label{fig:high_resolution velocity_high}
\end{figure}

\begin{figure}[htb!]
    \centering
    \subfloat[]{\includegraphics[width=1.0\textwidth]{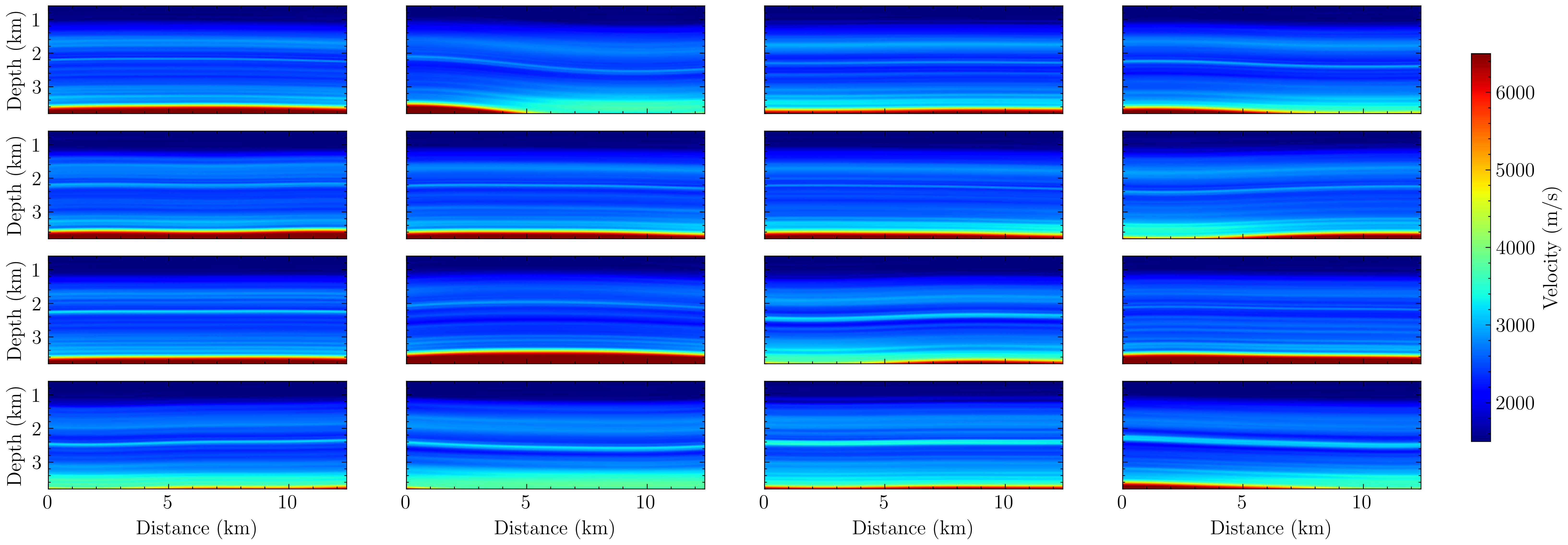}}\
    \subfloat[]{\includegraphics[width=0.5\textwidth]{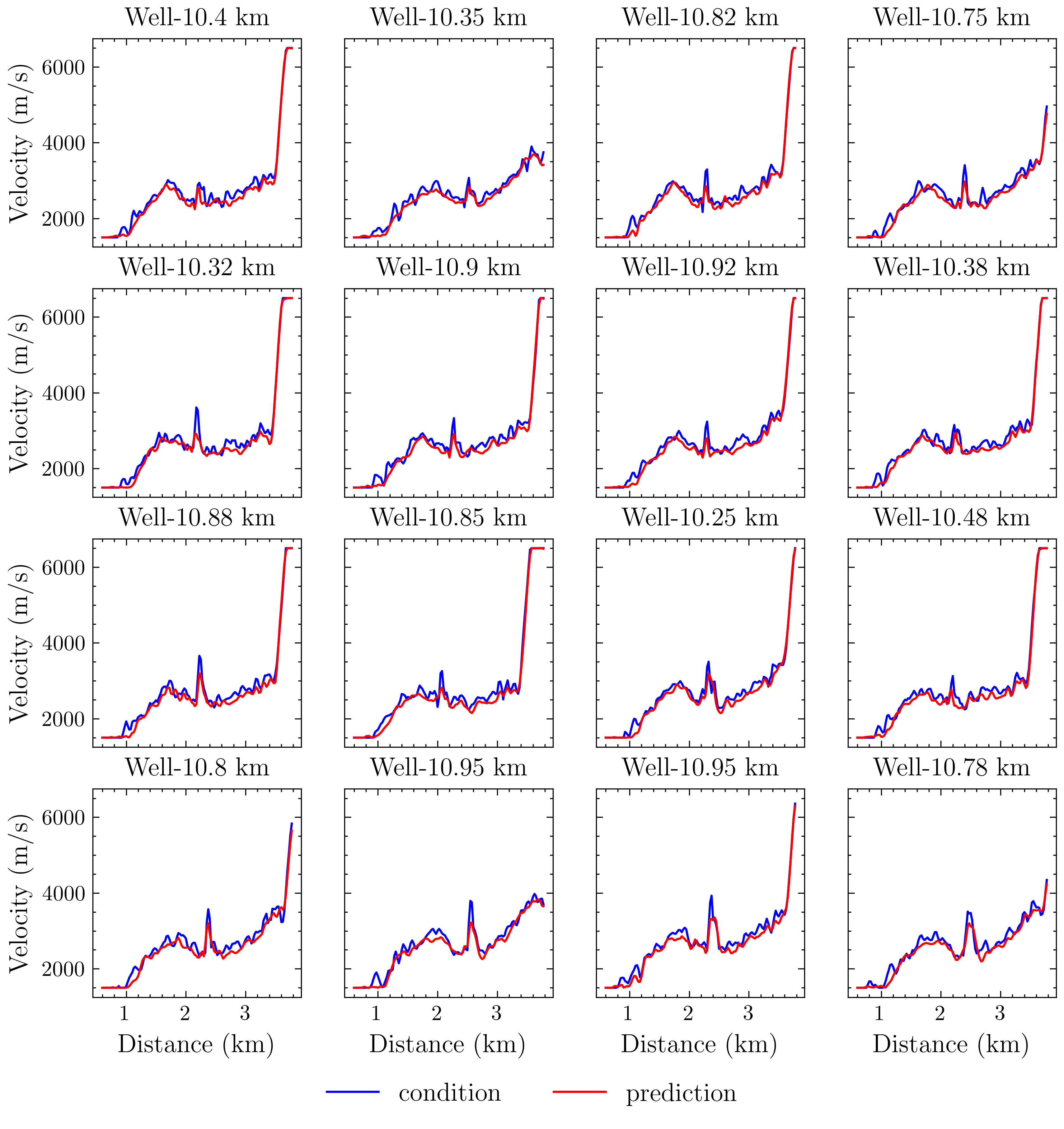}}    
    \caption{(a) 16 samples of the velocity using conditional diffusion model, and (b) the corresponding comparison between the extracted well profile from generated velocity models and given conditions.}
    \label{fig:well_training_high}
\end{figure}

Regarding the training velocity models, we generate 20000 samples with a resolution of $500\times128$ guided by the well logs and structural prior \citep{ovcharenko2022multi} (remove the water layer), and some samples of the training dataset are shown in Figure~\ref{fig:high_resolution velocity_high}. 
We use the same configuration as the OpenFWI test to train the conditional diffusion model, except we increase the size of the embedding layer for the well log embedding as we hope to incorporate high-resolution information in the wells. Figure~\ref{fig:well_training_high} shows 16 samples obtained by conditional diffusion models given different well-log profiles.
The matching between the well-log profiles and profiles extracted from the predicted velocity models is not perfect.
This drawback is coming from the challenge of the diffusion model to handle high-resolution images \citep{rombach2022high}. 
Compared to the training samples (Figure~\ref{fig:high_resolution velocity_high}), although the samples (Figrue~\ref{fig:well_training_high}a) fit the structure information and velocity changes, from the comparison of the well profiles, the prediction only matches the trend of the well log.

\begin{figure}[htb!]
    \centering
    {\includegraphics[width=0.68\textwidth]{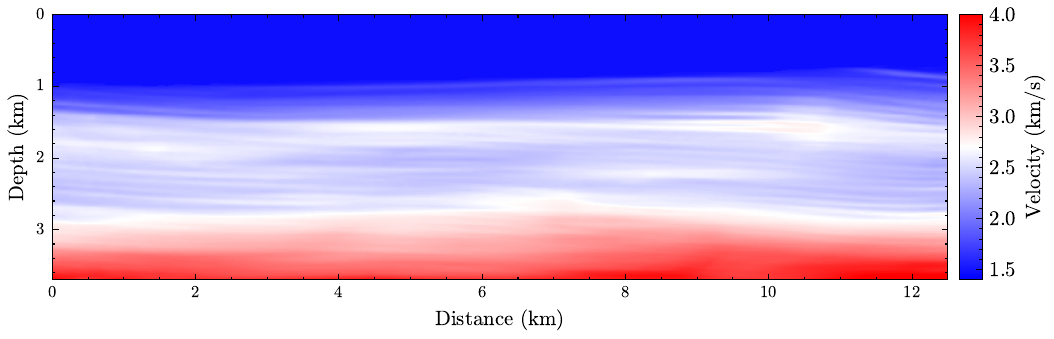}}\label{fig:cdiffusionFWI_high}   
    \caption{The inversion result from well-conditioned diffusion FWI where the conditional factor for well log is 9.}
    \label{fig:inversion_test_high}
\end{figure}

\begin{figure}[htb!]
    \centering
    \includegraphics[width=0.75\textwidth]{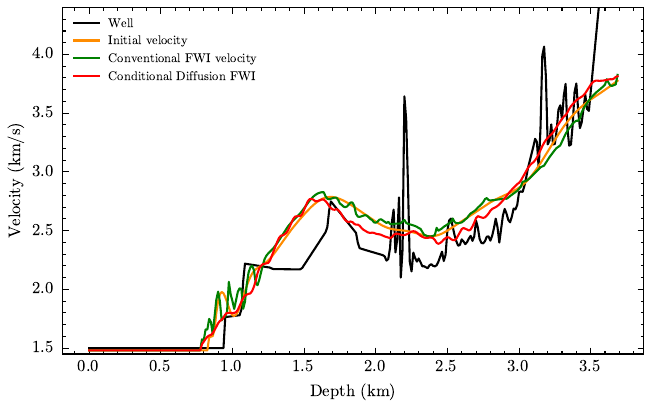}
    \caption{The vertical profiles comparison, at a distance of 10.5$km$, of the initial velocity model, and the inverted results using the conventional FWI, diffusion regularized FWI, and conditional diffusion regularized FWI. The proposed well-conditioned diffusion FWI matches the trend of the well better.}
    \label{fig:well_comp_high}
\end{figure}

Then, we further apply the proposed method to invert for the velocity, shown in Figure~\ref{fig:inversion_test_high}.
We observed the influence of the well condition on the inverted velocity even though the region is far away from the well log location, especially at the left part of the velocity model, whose wavenumber is higher than the conventional one (Figure~\ref{fig:inversion_test}b) and has velocity consistency with the right parts. 
We also extract the profiles at the well location of 10.5 $km$ for comparison, shown in Figure~\ref{fig:well_comp_high}.
Although the proposed method with a high-resolution diffusion model does not perform better than the one using a patch-based strategy, it is closer to the well log than the one obtained using the conventional FWI and matches the trend of the well log. 
To further enhance the proposed method with a diffusion model, we argue that utilizing a latent diffusion model \citep{rombach2022high} is beneficial. 
It can reduce computational cost while preserving high-quality generative capabilities. 
More importantly, it offers a structured latent space where disentanglement of features (e.g., geological variations) may be easier to achieve. 
We leave this extension for future investigation.

Also, for the field test in this study, unlike the synthetic tests, in which the well profile extracted from the conditional diffusion regularized FWI matched the well log condition, there is still a gap.
Beyond the bias of the training velocity itself, there is another reason for that, which is the presence of anisotropy in this region.
Hence, the well-log velocity (representing the velocity in the vertical direction) will not match the velocity inverted from an acoustic FWI \citep{sun2023anisotropic}. 
This causes the inherent gap between the inverted result and the well log, and it would be better to replace the modeling engine of the FWI with an anisotropic acoustic modeling to improve the inverted result of our proposed method.

\section{Conclusions}
\label{conclude}
We proposed a novel and promising way to include multi-modal prior information into FWI for subsurface reconstruction. We specifically utilized a conditional diffusion model for controllable velocity generation, in which multi-modal prior information can be feasibly incorporated, and combine it with full waveform inversion using a diffusion regularized FWI framework, yielding a method of multi-modal prior assisted regularized FWI. It enables the inverted results to satisfy the data, velocity distribution as well as other additional priors. During the training of the conditional diffusion model, the distribution of the velocity models and the multi-modal priors are stored in the neural networks and can be accessed by means of classifier-free guidance.
At the inversion stage, the guidance from the conditional diffusion model can help FWI improve the inverted results to match the multi-modal prior information.
This method allows for the feasible integration of the expertise from geologists into the FWI process in an automatic way.
Specifically, we test the conditional diffusion model-guided FWI approach using geological classes and well priors. 
The experiments on the OpenFWI dataset demonstrate the better performance of the proposed approach compared to conventional FWI and regular diffusion FWI. 
A further test on the field application also validates the effectiveness of the proposed method. 

\section{Acknowledgments}
We thank KAUST and the DeepWave Consortium sponsors for their support. This work utilized the resources of the Supercomputing Laboratory at KAUST, and we are grateful for that.

\bibliographystyle{plainnat}
\bibliography{example}
\end{document}